\documentclass[twocolumn, tighten, twocolappendix]{aastex631_mod}
\usepackage[T1]{fontenc}
\usepackage{microtype}
\usepackage{newtxtext}
\usepackage[varvw]{newtxmath}
\usepackage{mathtools}
\usepackage{enumitem}

\usepackage{booktabs}
\usepackage{multirow}
\usepackage{tikz}

\usetikzlibrary{shapes}
\hypersetup{pdfauthor={Errani et al.},
            pdftitle={},
            pdfkeywords={},
            bookmarksnumbered=true}

\newcommand{\rmx}{r_\mathrm{mx}}

\newcommand{\Vmx}{V_\mathrm{mx}}
\newcommand{\Mmx}{M_\mathrm{mx}}
\newcommand{\Tmx}{T_\mathrm{mx}}

\newcommand{\kms}{\mathrm{km\,s^{-1}}}
\newcommand{\Rh}{R_\mathrm{h}}

\newcommand{\rh}{r_\mathrm{h}}

\newcommand{\diff}{\mathrm{d}}
\newcommand{\Msol}{\mathrm{M_{\odot}}}
\newcommand{\Lsol}{\mathrm{L_{\odot}}}

\newcommand{\rs}{r_\mathrm{s}}
\newcommand{\Vc}{V_\mathrm{c}}

\newcommand{\kpc}{\mathrm{kpc}}
\newcommand{\pc}{\mathrm{pc}}
\newcommand{\Gyr}{\mathrm{Gyr}}
\newcommand{\rperi}{r_\mathrm{peri}}
\newcommand{\Tperi}{T_\mathrm{peri}}

\newcommand{\rapo}{r_\mathrm{apo}}
\newcommand{\mxzero}{_\mathrm{mx0}}
\newcommand{\dex}{\mathrm{dex}}
\newcommand{\rt}{{\,:\,}}
\newcommand{\sigmalos}{\sigma_\mathrm{los}}

\newcommand{\g}{\mathrm{g}}
\newcommand{\E}{\mathcal{E}}
\newcommand{\plus}[1] {^{\mathmakebox[\widthof{$^-$}][c]{+}#1}}
\newcommand{\minus}[1]{_{\mathmakebox[\widthof{$^-$}][c]{-}#1}}

\newcommand{\gnuplotline}[2][very thick]{\tikz[baseline={([yshift=-.5ex]current bounding box.base)}]{ \draw[#2,#1, line cap=round] (0,0) -- (0.3,0); }}
\newcommand{\gnuplotarea}[1]{\tikz[baseline={([yshift=-0.1ex]current bounding box.base)}]{ \path[#1,inner sep=0pt, line width=0.02cm, line cap=round] (0,0) rectangle (0.5,0.12); }}
\newcommand{\gnuplottriangle}[1][fill=black]{\tikz[baseline={([yshift=-0.7ex]char.base)}]{ \node[#1,regular polygon ,regular polygon sides=3, draw, inner sep=0.9pt, style={rotate=180}, rounded corners=0.2mm] (char) {};}}
\newcommand{\gnuplotcircle}[1][fill=black]{\tikz[baseline={([yshift=-0.6ex]char.base)}]{ \node[#1,shape=circle, draw, inner sep=1.3pt] (char) {};}}
\newcommand{\gnuplotcross}[1][draw=black]{\tikz[baseline={([yshift=-0.6ex]char.base)}]{ \node[#1, cross out, draw, inner sep=0.8pt, line width=0.05cm, line cap=round] (char) {};}}
\newcommand{\gnuplotcrosstwocolors}{\tikz[baseline={([yshift=-0.6ex]char.base)}]{ \node[fill=black!33, draw=black!33, shape=rectangle, draw, inner sep=2.4pt, rounded corners=0.2mm] (char) {};  \node[draw=GCcolor, cross out, draw, inner sep=0.8pt, line width=0.04cm, line cap=round] (char) {};    }}

\definecolor{newred}{HTML}{D73027}
\definecolor{newdarkblue}{HTML}{315e99}

\definecolor{newlightblue}{HTML}{74ADD1}    
\definecolor{orchid4}{HTML}{804080}   
\definecolor{steelblue}{HTML}{306080}   
\definecolor{darkmagenta}{HTML}{c000ff}
\definecolor{orangered}{HTML}{ff4500}

\definecolor{dwarfcolor}{HTML}{87ceeb}
\definecolor{GCcolor}{HTML}{ffd700}

\definecolor{purple}{HTML}{c080ff}
\definecolor{tan1}{HTML}{ffa040}
\definecolor{forestgreen}{HTML}{228b22}
\definecolor{skyblue}{HTML}{87ceeb}

\graphicspath{{./}{./Figures/}}

\begin{document}
\shorttitle{Micro galaxies in LCDM}
\shortauthors{Errani et al.}
\title{Micro galaxies in LCDM}

\author{Rapha\"el Errani}
\affiliation{McWilliams Center for Cosmology, Department of Physics, Carnegie Mellon University, Pittsburgh, PA 15213, USA}
\affiliation{Universit\'e de Strasbourg, CNRS, Observatoire Astronomique de Strasbourg, UMR 7550, F-67000 Strasbourg, France}
\email{errani@cmu.edu}

\author{Rodrigo Ibata}
\affiliation{Universit\'e de Strasbourg, CNRS, Observatoire Astronomique de Strasbourg, UMR 7550, F-67000 Strasbourg, France}

\author{Julio F. Navarro}
\affiliation{Department of Physics and Astronomy, University of Victoria, Victoria, BC V8P 5C2, Canada}

\author{Jorge Pe\~narrubia}
\affiliation{Institute for Astronomy, University of Edinburgh, Royal Observatory, Blackford Hill, Edinburgh EH9 3HJ, UK}

\author{Matthew G. Walker}
\affiliation{McWilliams Center for Cosmology, Department of Physics, Carnegie Mellon University, Pittsburgh, PA 15213, USA}

\received{2023 November 24}
\revised{2024 March 18}
\accepted{2024 April 16}

\begin{abstract}
A fundamental prediction of the Lambda Cold Dark Matter (LCDM) cosmology is the centrally divergent cuspy density profile of dark matter haloes. Density cusps render CDM haloes resilient to tides, and protect dwarf galaxies embedded in them from full tidal disruption. The hierarchical assembly history of the Milky Way may therefore give rise to a population of ``micro galaxies''; i.e., heavily stripped remnants of early accreted satellites, which may reach arbitrarily low luminosity. Assuming that the progenitor systems are dark matter dominated, we use an empirical formalism for tidal stripping to predict the evolution of the luminosity, size, and velocity dispersion of such remnants, tracing their tidal evolution across multiple orders of magnitude in mass and size. The evolutionary tracks depend sensitively on the progenitor distribution of stellar binding energies. We explore three cases that likely bracket most realistic models of dwarf galaxies: one where the energy distribution of the most tightly bound stars follows that of the dark matter, and two where stars are defined by either an exponential density or surface brightness profile. The tidal evolution in the size--velocity dispersion plane is quite similar for these three models, although their remnants may differ widely in luminosity. Micro galaxies are therefore best distinguished from globular clusters by the presence of dark matter; either directly, by measuring their velocity dispersion, or indirectly, by examining their tidal resilience. Our work highlights the need for further theoretical and observational constraints on the stellar energy distribution in dwarf galaxies.
\end{abstract}
\keywords{Cold dark matter (265); Dwarf spheroidal galaxies (420); Low surface brightness galaxies (940); Milky Way Galaxy (1054); N-body simulations (1083); Star clusters (1567); Tidal disruption (1696)}
\defcitealias{ENIP2022}{E+22}
\defcitealias{EN21}{EN21}
\defcitealias{EP20}{EP20}

\section{Introduction}
In the Lambda Cold Dark Matter (LCDM) cosmology, galaxies are predicted to form in the potential wells of dark matter overdensities \citep{WhiteRees1978}. Through a history of accretion and merger events, these overdensities give rise to a complex clustering hierarchy of \textit{haloes} and \textit{subhaloes} \citep[for reviews, see, e.g.][]{FrenkWhite2012, ZavalaFrenk2019}. 

Computer simulations suggest that the internal mass distribution of haloes is well approximated by the Navarro-Frenk-White (NFW) profile \citep{Navarro1996a,Navarro1997}. This means that CDM haloes are predicted to have remarkably high central densities: for NFW profiles, the density formally diverges as $\diff \ln \rho_\mathrm{NFW} / \diff \ln r \rightarrow -1$ for $r \rightarrow 0$. In contrast, some other theories of dark matter, such as self-interacting dark matter (SIDM), predict haloes with constant-density cores, i.e., $\diff \ln \rho_\mathrm{core} / \diff \ln r \rightarrow 0$ for $r \rightarrow 0$ \citep{Burkert2000, Spergel2000, Colin2002}.

Dwarf spheroidal (dSph) galaxies are promising observational probes of the properties of galactic dark matter subhaloes: their dynamical mass-to-light ratios, inferred from stellar kinematics, suggest that most dSphs are dark matter-dominated objects (\citealt{Mateo1993}, \citealt{Walker2007bLetter} \citealt{McConnachie2012}; for reviews, see \citealt{Mateo1998}, \citealt{Simon2019Review}, \citealt{Battaglia2022}). 

For the Fornax and Sculptor dSphs, several studies find evidence for cored dark matter density profiles \citep{Walker2011, Amorisco2012, AmoriscoAgnelloEvans2013, Diakogiannis2017, Pascale2018, Read2019}. Instead, for the Draco dSph, most studies favor a cuspy density profile \citep{Jardel2013,Read2018,Massari2020,Hayashi2020}. Whether this ``diversity'' \citep{Oman2015} is due to the intrinsic properties of dark matter, or driven by baryonic effects \citep[see, e.g.][]{Santos-Santos2020}, remains a matter of debate. 

An independent approach to infer properties of dark matter substructures on galactic scales is to study their response to tides. Dwarf galaxies that have been accreted onto the Milky Way are subject to tidal forces, which induce mass loss and structural changes \citep{Penarrubia2008}. 
In the case of LCDM, the centrally divergent density profile of subhaloes renders them very resilient to the effects of tides. Indeed, numerical simulations suggest that smooth tidal fields do not fully disrupt NFW subhaloes (see, e.g., \citealt{Penarrubia2010}, \citealt{vdb2018} and \citealt{EP20}, hereafter \citetalias{EP20}) but rather lead them to asymptotically approach a stable remnant state (\citealt{EN21}, hereafter \citetalias{EN21}). 

On the other hand, for cored dark matter substructures, tides may trigger a runaway process that leads to their full disruption \citep{Penarrubia2010, ENPFI2023}. The mere existence of heavily stripped substructures can hence help to constrain the density structure of haloes and thereby the nature of dark matter. Crucially, stars embedded inside cuspy dark matter subhaloes will be protected from full tidal disruption if their distribution of binding energies within the subhalo extends all the way to the most-bound states (\citealt{ENIP2022}, hereafter \citetalias{ENIP2022}). Heavily stripped dwarf galaxies may thereby give rise to a population of ``micro galaxies'' (\citetalias{EP20}), i.e., co-moving groups of stars embedded in heavily stripped dark matter subhaloes. 

Detailed modeling of the structural changes that dwarf galaxies undergo when subject to strong tidal fields is complicated by the limited particle number and spatial resolution of $N$-body simulations: insufficient resolution leads to an inaccurate evolution of structural parameters (\citetalias{EP20}, \citetalias{EN21}) and may result in the artificial disruption of substructures \citep{vdb2018}. This is of particular relevance for dwarf galaxies that pass through the inner regions of the Milky Way, where the combined tidal field of disk, bulge, and halo is strongest, significantly increasing the rate of tidal stripping \citep{DOnghia10,EPLG17,Kelley2019}. 

In this work, we model the observable properties of the faintest Milky Way satellites, assuming that they are embedded in LCDM subhaloes, and accurately following their tidal evolution over many orders of magnitude in mass, size and luminosity. Our models use the empirical tidal stripping framework introduced in \citetalias{ENIP2022}, which allows us to make predictions for spatial and mass scales inaccessible to current cosmological simulations.  

We discuss our findings in the light of recently discovered Milky Way satellites \citep{Torrealba2019clusters, Mau2020, Cerny22, Cerny23, Smith2023} which have sizes and luminosities at the interface between the globular cluster and dwarf galaxy regime. Could these satellites have their origin in accreted dwarfs and be among the first micro galaxy candidates?

The paper is structured as follows. We show how resolution limits of cosmological $N$-body simulations affect the tidal survival of DM subhaloes in Section~\ref{sec:arti_disrupt}. In Section~\ref{sec:tidal_evo_model}, we summarize the empirical tidal stripping framework used in this study, and apply it to model the tidal evolution of dwarf galaxies over many orders of magnitude in mass loss. 
In Section~\ref{sec:application_to_data} we compare the model predictions against observed properties of Milky Way satellites, paying particular attention to systems at the interface of the globular cluster and dwarf galaxy regime. 
Finally, we summarize our main results and conclusions in Section~\ref{sec:conclusions}.

\section{Artificial Tidal Disruption in Cosmological \textit{N}-body Simulations}
\label{sec:arti_disrupt}
Convergence in $N$-body simulations depends on a combination of the spatial resolution (force softening length $\epsilon$, or particle mesh cell size $\Delta x$), particle number $N$ and time step $\Delta t$ (see, e.g., \citealt{Power2003}, \citealt{Springel2008}, \citealt{vdBOgiya2018}, \citetalias{EN21}). In this section, we perform controlled simulations to illustrate that the resolution of current cosmological $N$-body simulations is not enough to reliably trace the tidal evolution of most subhaloes that pass through the inner regions of the Milky Way. 

\subsection{Pericentric/Apocentric Radii and Accretion Redshift}
\label{subsec:raporperi}

\begin{figure}
 \begin{center}
 \includegraphics[width=\columnwidth]{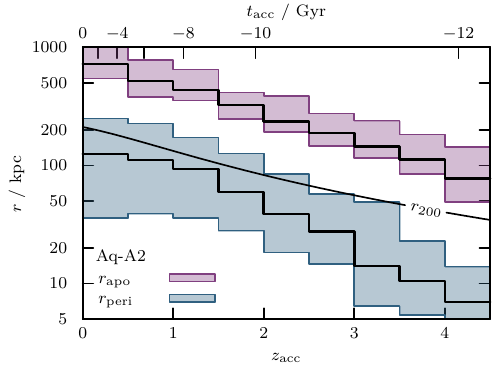}
\end{center}
\caption{Distribution of apocentre $r_\mathrm{apo}$~(\protect\gnuplotarea{draw=orchid4, fill=orchid4!35}) and pericentre $r_\mathrm{peri}$~(\protect\gnuplotarea{draw=steelblue, fill=steelblue!35}) for subhaloes in a Milky Way--like host halo as a function of accretion redshift $z_\mathrm{acc}$ (assuming Aquarius cosmological parameters, $H_0=73\,\kms \mathrm{Mpc}^{-1}, \Omega_\mathrm{m}=0.25, \Omega_\Lambda=0.75$). The distributions are computed from subhaloes with peak virial masses of $M_\mathrm{200} \geq 10^8\,\Msol$ in the Aquarius-A2 simulation. Median radii are shown as solid lines, with shaded regions corresponding to the $16^\mathrm{th}-84^\mathrm{th}$ percentiles.
On average, subhaloes accreted at earlier $z_\mathrm{acc}$ have smaller apo- and pericentres than those accreted more recently. Subhaloes with $r_\mathrm{peri} < 20\,\kpc$ have $z_\mathrm{acc} \gtrsim 2$ (i.e., they were accreted $\gtrsim 10\,\Gyr$ ago, see top axis). For reference, the evolution of the host halo virial radius $r_{200}$ is shown as a black solid curve.}
\label{fig:accretion_history}
\end{figure}

Cosmological simulations indicate that the present-day pericentric and apocentric radii of subhaloes correlate strongly with the time when they were first accreted onto the main halo. Figure~\ref{fig:accretion_history} illustrates this result for the case of the Aquarius-A2 simulation \citep{Springel2008}, where we have tracked the orbits of all subhaloes with virial\footnote{The virial radius $r_{200}$ encloses a mass $M_{200}$ so that the mean density $M_{200}/ (4/3~ \pi r_{200}^3)$ is $200\times$ larger than the critical density for closure, $\rho_\mathrm{crit}$. At redshift $z=0$, $\rho_\mathrm{crit}=3 H_0^2 / (8\pi G)$.\label{footnote:virial}} mass $M_{200} \gtrsim 10^8\,\Msol$ accreted onto the A2 main halo (as in \citealt{EPLG17}). Subhalo orbits have been computed as point-masses in a time-evolving analytical potential fitted to the main halo (eq.~22 and 23 in \citealt{buisthelmi16}). 
This setup allows us to follow the orbits of all subhaloes resolved at accretion until $z = 0$, without losing subhaloes to artificial disruption. 

The integration includes the adiabatic orbital contraction in the growing main halo potential, but neglects the effects of dynamical friction, which generally would affect only the few most massive satellites \citep{PenarrubiaBenson2005}, reducing their orbital radii even further. Clearly, subhaloes that orbit the inner regions of the main halo have been on average accreted earlier and have been therefore subjected to tidal effects for longer. 

Our results show\footnote{Comparison against the orbits of Milky Way (MW) satellite galaxies \citep{Li2021_Gaia, Battaglia2022_Gaia} suggests that the subhaloes in our controlled Aq-A2 setup are on average on more radial orbits than MW satellites. This resembles the ``tangential velocity excess'' of MW satellites discussed in \citet{Cautun2017}. Note, however, that our setup assumes a spherically symmetric halo and does not include a disc, which may play an important role in shaping the orbital anisotropies of satellites in the inner regions of our Galaxy \citep[see][]{Riley2019}.} that, on average, subhaloes on orbits with pericentres $\rperi\lesssim10\,\kpc$ were accreted more than $\gtrsim 10\,\Gyr$ ago. A subhalo on an orbit with $\rperi=10\,\kpc$ and $\rapo = 50\,\kpc$ has completed more than $\gtrsim 15$ orbital periods since accretion; subhaloes in the inner regions of the Milky Way have been subjected to strong tidal fields for extended periods of time. 

\subsection{Tidal Disruption and Numerical Resolution}
To illustrate the effects of insufficient numerical resolution on tidal evolution, we evolve four different $N$-body realizations of an NFW subhalo in a static isothermal host potential with a circular velocity $V_\mathrm{host} = 220\,\kms$ (see Eq.~\ref{eq:host_potential}). The $N$-body realizations differ only in the number of $N$-body particles used to simulate the subhalo.

\subsubsection{Initial Conditions}
As an example, we consider an NFW subhalo with a virial mass of $M_{200} \approx 7 \times 10^9\,\Msol$, close to the redshift $z=0$ hydrogen cooling limit \citep{Benitez-Llambay2020}. 
For an average concentration of $c=r_{200}/r_\mathrm{s} \approx 13$ \citep{Ludlow2016}, this corresponds to an initial characteristic mass and size of $M\mxzero \equiv M(<r\mxzero) = 2\times10^9\,\Msol$ and $r\mxzero = 7\,\kpc$, respectively (see Eq.~\ref{eq:NFW} for a definition). 

Using the code introduced in \citetalias{EP20}, available online\footnote{The code used to generate $N$-body models as described in \citetalias{EP20}, as well as an implementation of the \citetalias{EN21} tidal evolution model are available at \url{https://github.com/rerrani}.\label{footnote:github}}, we generate four $N$-body realizations of this NFW subhalo: For the first realization, labelled \texttt{A1}, we choose an $N$-body particle mass identical to that used in the Aquarius-A1 simulation, $m_\mathrm{p} = 1.71\times10^3\,\Msol$. Similarly, for the second realization, labelled \texttt{A2}, we use $m_\mathrm{p} = 1.37\times10^4\,\Msol$, matching the particle mass of the Aquarius-A2 simulation. The models \texttt{A3} and \texttt{A4} adopt particle masses that match those of the Aquarius-A3 and A4 resolution levels, respectively. The parameters of these simulations are summarized in Table~\ref{tab:parameter_overview}. As the NFW profile has a divergent total mass, we taper the profile exponentially beyond $10\,\rs$ (where by $\rs$ we denote the NFW scale radius as in Eq.~\ref{eq:NFW}). Consequently, the \texttt{A1}, \texttt{A2}, \texttt{A3} and \texttt{A4} realizations of our example subhalo consist of $N=4.4\times10^6$, $5.6\times10^5$, $1.5 \times 10^5$ and $1.9 \times 10^4$ particles, respectively. 
\begin{table}
\centering
\caption{Parameters of the example subhalo model considered in the $N$-body experiments of Sec.~\ref{sec:arti_disrupt} and the empirical models of Sec.~\ref{sec:tidal_evo_model}. The initial subhalo mass $M\mxzero$ is chosen to approximately match the (redshift $z=0$) hydrogen cooling limit \citep{Benitez-Llambay2019}, and the initial subhalo characteristic size $r\mxzero$ is that of a $z=0$ average-concentration halo \citep{Ludlow2016}. Listed also are the host halo and orbit properties underlying the $N$-body and empirical models.}
\label{tab:parameter_overview}
\begin{tabular}{l@{\hskip 0.8cm} l @{\hskip 0.8cm}l}
\toprule
\multirow{7}{*}{\rotatebox[origin=t]{90}{\textsc{S u b h a l o}$~~$}}
 & Profile          & NFW (Eq.~\ref{eq:NFW}) \\
 & $M\mxzero$, $V\mxzero$, $r\mxzero$       & {$2\times10^9\, \Msol$, $35\,\kms$, $7\,\kpc$}  \\[0.3cm]
 & $N$-body $\Delta x$       & {$r\mxzero/256 \approx 27\,\pc$}\\
 & \multirow{1}{*}{$N$-body $m_\mathrm{p}$}      & \texttt{A1}: {$1.71\times10^3\,\Msol$}\\ 
 &                                               & \texttt{A2}: {$1.37\times10^4\,\Msol$}\\ 
 &                                               & \texttt{A3}: {$4.91\times10^4\,\Msol$}\\ 
 &                                               & \texttt{A4}: {$3.93\times10^5\,\Msol$}\\ \midrule
\multirow{4}{*}{\rotatebox[origin=t]{90}{\textsc{H o s t}}}
& Profile          & Isothermal (Eq.~\ref{eq:host_potential}) \\
& $V_\mathrm{host}$& {$220\,\kms$} \\
& $\rperi$         & {$10\,\kpc$} \\
& $\rperi\rt\rapo$ & {$1\rt5$}\\

\bottomrule
\end{tabular}
\end{table}

\subsubsection{Simulation Code}
The simulations are carried out using the particle-mesh code \textsc{Superbox} \citep{Fellhauer2000}. The code employs a high-resolution cubic grid co-moving with the subhalo, with a cell size of $\Delta x \approx r\mxzero/256 \approx 27\,\pc$. \textsc{Superbox} uses two additional grids, one co-moving of medium resolution ($10\,r\mxzero/256$), and a fixed low-resolution ($500\,\kpc/256$) grid containing the entire simulation box.
For reference, the force softening length of the Aquarius-A1 simulation equals $\epsilon=22\,\pc$ \citep{Springel2008}, similar to the particle-mesh cell size $\Delta x \approx 27\,\pc$ of the highest-resolving grid used in our numerical experiments. Note that the only parameter varied in the simulations discussed in this section is the $N$-body particle mass $m_\mathrm{p}$, and thereby the total number of $N$-body particles in each simulation.

\begin{figure}
 \begin{center}
  \includegraphics[width=\columnwidth]{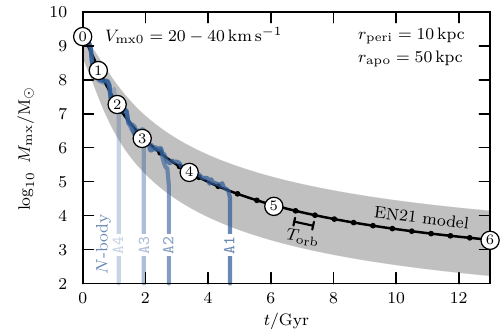}
\end{center}
\caption{Tidal stripping of a subhalo with an initial mass of $M\mxzero = 2\times10^9\,\Msol$ in a Milky Way--like host (see Table~\ref{tab:parameter_overview}), computed using $N$-body simulations (blue curves) and the empirical \citetalias{EN21} model (black solid curve). The four $N$-body simulations have particle masses $m_\mathrm{p}$ chosen to match those of the Aquarius-A1\,(\protect\gnuplotline[line width=0.08cm]{newdarkblue, opacity=0.73}), A2\,(\protect\gnuplotline[line width=0.08cm]{newdarkblue, opacity=0.6}), A3\,(\protect\gnuplotline[line width=0.08cm]{newdarkblue, opacity=0.4}) and A4\,(\protect\gnuplotline[line width=0.08cm]{newdarkblue, opacity=0.265}) resolution levels, respectively. The evolution of the subhalo mass $\Mmx$ is shown as a function of time $t$. Numbered circles $\protect\textcircled{0},\,...\,,\protect\textcircled{6}$ mark remnant masses of $\log_{10}\Mmx/M\mxzero=0,\,-1,\,...\,,\,-6$. Shaded regions correspond to initial conditions spanning a range of circular velocities $20\leq V\mxzero/\kms \leq 40$. Insufficient numerical resolution causes the $N$-body models disrupt \textit{artificially}, whereas the empirical model predicts an evolution towards a stable asymptotic remnant state.}
\label{fig:nbody_vs_model}
\end{figure}

\subsubsection{Simulation Results}
\label{Sec:NbodyResults}
The NFW subhalo models are placed on an orbit with pericentric and apocentric distance $\rperi = 10\,\kpc$ and $\rapo = 50\,\kpc$, respectively, and evolved for $13\,\Gyr$. Figure~\ref{fig:nbody_vs_model} shows the evolution of $\Mmx$, the bound\footnote{The bound mass is computed by (1) defining the subhalo centre through the shrinking spheres method \citep{Power2003}, (2) computing the subhalo potential under the assumption of spherical symmetry, (3) removing particles un-bound in this potential. These steps are iterated until convergence.} mass within the radius, $\rmx$, where the circular velocity peaks, as a function of time. The four $N$-body realizations are shown in different shades of blue. Over the first two decades in mass loss, all four $N$-body realizations show a similar mass evolution. Beyond that point, however, the evolution differs between the models. The \texttt{A4} model fully disrupts after only $\sim1\,\Gyr$; the \texttt{A2} model fully disrupts after $\sim 2.5\,\Gyr$. The \texttt{A1} model survives slightly longer, but eventually also disrupts after a total of $\sim 3.5\,\Gyr$ of evolution. 

The increase in $N$-body particle number has hence delayed, but not prevented, the full disruption of the subhalo. This numerical experiment suggests that even simulations with a resolution as high as that of the Aquarius-A1 simulation will suffer from the artificial depletion of substructures on orbits that reach the innermost regions of the Galaxy. Even before the subhalo is (artificially) disrupted, its structural evolution is compromised by insufficient resolution: the convergence study listed in appendix~A of \citetalias{EN21} shows that, once the number of bound particles within $\rmx$ drops below $N_\mathrm{mx} <3000$, resolution artifacts result in subhalo densities being systematically underestimated. 
Evidently, a tool different from classical cosmological $N$-body simulations is needed to model the tidal remnants of heavily stripped dwarf galaxies.

To address this issue, we use an empirical model for the tidal evolution of subhaloes (and embedded dwarf galaxies), which allows us to study their tidal evolution over many orders of magnitude in mass loss. A black solid curve in Figure~\ref{fig:nbody_vs_model} shows the evolution of the subhalo computed using the empirical \citetalias{EN21} model, available online (see footnote~\ref{footnote:github}). The model suggests that mass loss keeps decelerating, and that a stable remnant state is asymptotically approached. The details of this model, as well as its extension to model the evolution of dwarf galaxies, are summarized in the next chapter.

\section{Tidal Evolution Model}
\label{sec:tidal_evo_model}
We summarize now the empirical model for tidal stripping used in this work. 
\subsection{Model for the Evolution of the Dark Matter Component}
\label{sec:DM_evo}
For the evolution of the dark matter component, we rely on the empirical model introduced and tested in \citetalias{EN21}. In the initial conditions, the subhalo is assumed to follow an NFW density profile with a scale radius $r_\mathrm{s}$ and a scale density $\rho_\mathrm{s}$, 
\begin{equation}
\label{eq:NFW}
 \rho_\mathrm{NFW}(r) =  \rho_\mathrm{s}  \left(r/r_\mathrm{s} \right)^{-1}   \left(1+r/r_\mathrm{s}\right)^{-2} ~. 
\end{equation}
Instead of referring to $r_\mathrm{s}$ and $\rho_\mathrm{s}$ directly, we characterize the subhalo using two equivalent parameters: the peak velocity $\Vmx\approx 1.65\, \rs \sqrt{G\rho_\mathrm{s}}$ of the subhalo circular velocity profile $\Vc = \sqrt{GM(<r)/r}$, and the radius $\rmx \approx 2.16\, \rs$ where the peak velocity is reached. 
The effect of tides on the subhalo is modeled as an exponential truncation and a renormalization of the density profile (see \citetalias{EN21} equation 7),
\begin{equation}
 \rho(r) = \rho_\mathrm{NFW}(r) ~ \times ~ \exp(-r/r_\mathrm{cut}) / (1+r_\mathrm{s}/r_\mathrm{cut})^\kappa 
\end{equation}
with $\kappa \approx 0.3$ obtained from fits to $N$-body simulations. For $r_\mathrm{cut}/r_\mathrm{s} \rightarrow \infty$, the above equation reduces to the initial NFW profile, whereas for strong tidal truncation, $r_\mathrm{cut}/r_\mathrm{s} \rightarrow 0$, the profile converges to an exponentially truncated cusp,
\begin{equation}
 \rho_\mathrm{asy}(r) = \rho_\mathrm{cut} \exp(-r/r_\mathrm{cut}) (r/r_\mathrm{cut})^{-1} ~.
 \label{eq:rho_asy}
\end{equation}
For the exponentially truncated cusp, $\rmx \approx 1.79\, r_\mathrm{cut}$, $\Vmx \approx 1.94\,  r_\mathrm{cut} \sqrt{ G \rho_\mathrm{cut} }$. The truncated cusp has a total mass of $M_\mathrm{cut} = 4 \pi \rho_\mathrm{cut} r_\mathrm{cut}^3$, roughly twice as large as the mass $\Mmx\approx 0.54\,M_\mathrm{cut} $ enclosed within $\rmx$.
As tides strip the subhalo, its characteristic velocity $\Vmx$ and size $\rmx$ decrease, following a \textit{tidal track} (see \citealt{Penarrubia2008}) that couples the evolution of $\Vmx$ to that of $\rmx$ relative to their initial values (here, we use the track of \citetalias{EN21}; see their equation~5):
\begin{equation}
 \Vmx / V\mxzero = 2^\alpha (\rmx / r\mxzero)^\beta [1 + (\rmx/r\mxzero)^2]^{-\alpha}
 \label{eq:Eqn21track}
\end{equation}
with $\alpha \approx 0.4$ and $\beta \approx 0.65$ obtained from fits to $N$-body simulations.

We also use the \citetalias{EN21} model to estimate the time it takes for a subhalo to be stripped to a given remnant mass, assuming that the subhalo is on an orbit with a constant pericentre $\rperi$ and apocentre $\rapo$ within an isothermal host potential,
\begin{equation}
 \label{eq:host_potential}
 \Phi_\mathrm{host} = V_\mathrm{host}^2 ~ \ln \left( r / r_0 \right)~,
\end{equation}
where by $V_\mathrm{host}$ we denote the host's constant circular velocity, and $r_0$ is an arbitrary scale radius. 

The rate of tidal stripping depends on the density contrast between subhalo and host, measured at the pericentre (\citetalias{EN21} equations 12 and 15). For a given pericentre, the only effect of orbital eccentricity is to \textit{delay} tidal evolution with respect to the evolution on a circular orbit. For subhaloes that are underdense with respect to the mean host density at pericentre, the tidal evolution gradually decelerates until a final remnant state is reached where the subhalo mean density within $\rmx$ is $\approx 16$ times higher than the mean host density at the pericentre. In terms of time scales, heavily stripped subhaloes converge towards a characteristic circular time $ \Tmx = 2 \pi \rmx / \Vmx  \approx \Tperi/4$ set by the circular time of the host at the pericentre $ \Tperi = 2 \pi \rperi / V_\mathrm{host} $.

As an example, we use the empirical method outlined here to model the tidal evolution of a dark matter subhalo with the same characteristic mass and size as in the $N$-body example of Sec.~\ref{Sec:NbodyResults} (see Tab.~\ref{tab:parameter_overview}), and place it on the same orbit. The time evolution of the dark matter subhalo is shown in Fig.~\ref{fig:nbody_vs_model}: for the range of remnant masses resolved in the $N$-body simulations, there is excellent agreement between the empirical model and the simulation results. 
During $\sim13\,\Gyr$ of evolution, the subhalo is stripped to a remnant mass $\sim10^6$ times smaller than its initial mass. 

\begin{figure}
 \begin{center} 
  \includegraphics[width=\columnwidth]{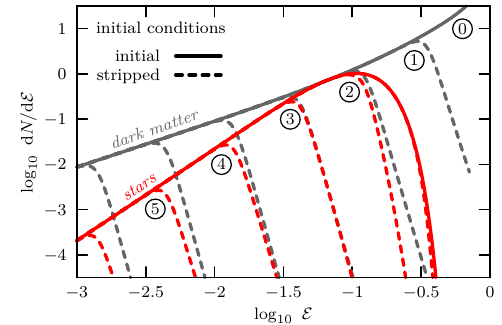}
\end{center}
\caption{Solid curves show the initial energy distribution $\diff N / \diff \E$ of dark matter (grey \protect\gnuplotline{gray}) and stars (red \protect\gnuplotline{red}), corresponding to snapshot~$\protect\textcircled{0}$ shown in Fig.~\ref{fig:2D_snapshots}. Dashed curves show the tidally truncated energy distributions of dark matter~(\protect\gnuplotline{gray,dashed}) and stars~(\protect\gnuplotline{red,dashed}) in the initial conditions for remnant subhalo masses of $\log_{10}\Mmx/M\mxzero=0,\,-1,\,...\,,\,-5$ (snapshots $\protect\textcircled{1},\,...\,,\protect\textcircled{5}$ ). The stars initially follow a 2D exponential surface brightness profile (model \texttt{exp2D}, see Sec.~\ref{subsubsec:exp2D}) with a half-light radius of $R_\mathrm{h0} = r\mxzero/16$. The stellar energy distribution is normalized so that $\max(\diff N / \diff \E)=1$; the normalization for the dark matter energy distribution is arbitrary.}
\label{fig:stellar_dnde}
\end{figure}

\begin{figure*}
 \begin{center} 
  \includegraphics[width=\textwidth]{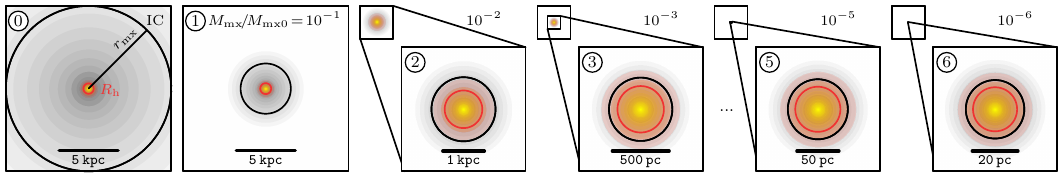}
\end{center}
\caption{Tidal evolution of a dwarf galaxy embedded in a dark matter subhalo, computed using the empirical energy-truncation model of \citetalias{ENIP2022}. The dark matter surface-density $\rho_\mathrm{2D}$ is shown in grey, and the surface-brightness $\Sigma_\star$ of the embedded dwarf is shown in colour. Snapshots, labelled $\protect\textcircled{0},\,...\,,\protect\textcircled{6}$, are shown for fixed subhalo remnant masses of $\log_{10}\Mmx/M\mxzero=0,\,-1,\,...\,,\,-6$. The scale is kept constant between the first two panels and the smaller top-left panels. The stars initially follow a (2D) exponential surface brightness profile with half-light radius $R_\mathrm{h0} = r\mxzero/16$ (model \texttt{exp2D}, see Sec.~\ref{subsubsec:exp2D}). Initially, the stars are hardly affected by tides, and mainly dark matter is lost (first three panels). Once the tidal energy truncation reaches the stellar component, the characteristic size of the dark matter halo $\rmx$ (shown as black circles) and the stellar half-light radius $\Rh$ (red circles) evolve in parallel (last three panels).}
\label{fig:2D_snapshots}
\end{figure*}

\subsection{Model for the Evolution of Stellar Tracers}
We now turn our attention to modelling the effect of tides on a stellar system embedded within a dark matter subhalo. We use the model outlined and tested in \citetalias{ENIP2022}, where stars are assumed to be massless and collisionless tracers within the subhalo potential. The effect of tides on a dwarf galaxy can then be described through subsequent truncations in the energy distributions of dark matter and stars. 

Under the assumptions of spherical symmetry and an isotropic velocity dispersion, for a given subhalo potential $\Phi(r)$, the initial stellar component is fully defined by its energy distribution
\begin{equation}
 \left. \diff N_\star / \diff \E \right|_{\mathrm{i}} = (4 \pi)^2~ f_\star\left(\E\right)~ p\left(\E\right)
\end{equation}
where by $f_\star(\E)$ we denote the initial stellar distribution function, and by $p\left(\E\right)$ we denote the density of states in the initial NFW potential. Both are functions of energy $\E = 1-E/\Phi_0$, where $E = v^2/2 + \Phi(r)$, and $\Phi_0$ denotes the potential minimum. For a given initial stellar density profile $\rho_\star(r)$, the stellar distribution function can be obtained by Eddington inversion. We compute $f_\star(\E)$ and $p\left(\E\right)$ using the implementation described in \citetalias{EP20}, available online (see footnote~\ref{footnote:github}). We will discuss our choices of initial stellar density profiles separately in Sec.~\ref{sec:cuspy_vs_cored_tracks}, and now proceed assuming that an appropriate initial $\diff N_\star / \diff \E$ has been defined.

The effect of tides on the stellar energy distribution $\diff N_\star / \diff \E$ can be approximated by a tapered truncation. We model the truncation empirically through (\citetalias{ENIP2022} eq.~9),
\begin{equation}
 \left. \diff N_\star / \diff \E \right|_{\mathrm{i,t}} = \frac{\left. \diff N_\star / \diff \E \right|_{\mathrm{i}}}{ 1 + \left(a \E / \E_\mathrm{mx,t} \right)^b}
\end{equation}
where by $ \left. \diff N_\star / \diff \E \right|_{\mathrm{i,t}}$ we denote the truncated energy distribution \textit{in the initial conditions}, and $\E_\mathrm{mx,t}$ is the energy scale beyond which the energy distribution is truncated. The parameters $a \approx 0.85$, $b \approx 12$ are obtained through fits to $N$-body simulations. 

As an example, Fig.~\ref{fig:stellar_dnde} shows the initial stellar and dark matter energy distributions for a stellar tracer with a 2D exponential surface brightness profile, deeply embedded in an NFW halo ($\Rh / \rmx = 1/16$). Energies $\E = 1 - E/\Phi_0$ are measured relative to the potential minimum $\Phi_0$. In this notation, the most-bound energy state (the ground state) is $\E = 0$, and the boundary between bound and unbound particles lies at $\E = 1$.

The energy-truncated system is initially out of equilibrium. We model the return to virial equilibrium empirically using fits to $N$-body simulations. The virialization process preserves (on average) the order of energies: the most-bound particles prior to virialization are also (on average) the most-bound particles after virialization. The mapping of \textit{initial} pre-virialization energies $\E$ to \textit{final} energies $\E_\mathrm{f}$ after virialization follows, on average, the empirical relation (\citetalias{ENIP2022} eq.~12)
\begin{equation}
 \bar{ \E}_\mathrm{f} = \left[ 1 + \left( c \E/\E_\mathrm{mx,t}  \right)^d  \right]^{1/d} 
\end{equation}
with $c \approx 0.8$ and $d \approx -3$. 
Note that the pre-virialization energies $\E$ are defined in the initial NFW potential, whereas the energies $\E_\mathrm{f}$ after virialization are defined in the tidally stripped subhalo potential.

Under the assumption of isotropic velocity dispersion and spherical symmetry, we then reconstruct the distribution function from the virialized energy distribution in the evolved subhalo potential. All observable properties of the stellar component can be derived from the evolved distribution function.

The simple empirical method outlined here allows us to model the tidal evolution of accreted substructures down to \textit{tiny} remnant masses and sizes. In Figure~\ref{fig:2D_snapshots}, we apply the empirical tidal stripping model to follow the evolution of a dwarf galaxy with a (2D) exponential surface brightness profile with an initial projected half-light radius of $\Rh \approx 440\,\pc$, and an initial luminosity of $L\approx10^6\,\Lsol$. We embed the dwarf galaxy in a subhalo with the same characteristic mass and size as in the $N$-body example of Sec.~\ref{Sec:NbodyResults} (see Tab.~\ref{tab:parameter_overview}). This results in a line-of-sight velocity dispersion of $\langle \sigmalos^2 \rangle^{1/2} \approx 12\,\kms$. The luminosity, size, and velocity dispersion of the dwarf galaxy model are therefore roughly comparable to those of classical Milky Way satellites like Ursa Minor, Sextans, or Sculptor. We model the tidal evolution of the dwarf galaxy assuming the same orbit as in Sec.~\ref{Sec:NbodyResults}.

After $\sim6\,\Gyr$ of tidal evolution, snapshot \textcircled{5}, the dwarf galaxy has been stripped to the size of a \textit{micro galaxy}, with an enclosed dark matter mass of $\sim10^4\,\Msol$, a luminosity of only $\sim50\,\Lsol$ and a size of several tens of parsecs, well beyond the resolution limits of $N$-body simulations like Aquarius-A1 (see Figure~\ref{fig:nbody_vs_model}). After $\sim12\,\Gyr$ of evolution, snapshot \textcircled{6}, the stellar system has been stripped to roughly solar luminosity, but still encloses $\sim10^3\,\Msol$ of dark matter.

\begin{figure*}
\centering
\hspace*{-1.0cm}\includegraphics[width=15.5cm]{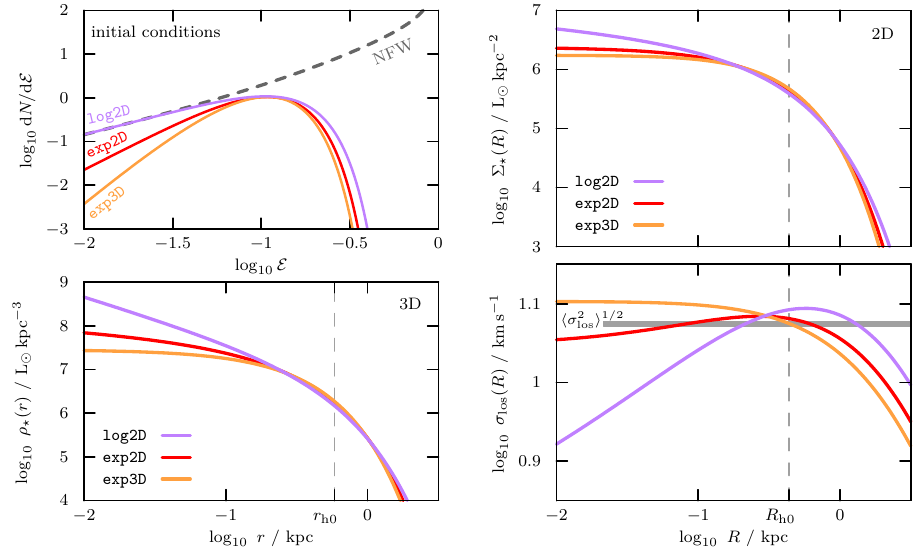}
\caption{\textit{Top left.} Initial energy distributions for the \texttt{exp2D} (\protect\gnuplotline{red}), \texttt{exp3D} (\protect\gnuplotline{tan1}) and \texttt{log2D} (\protect\gnuplotline{purple}) stellar models (Sec.~\ref{subsubsec:exp2D}, \ref{subsubsec:exp3D}, \ref{subsubsec:log2Dstars})  embedded in an NFW dark matter halo with parameters as in Tab~\ref{tab:parameter_overview}. All three models share the same projected half-light radius of $R_\mathrm{h0}/r\mxzero = 1/16$. For comparison, the energy distribution of the underlying NFW halo is shown as a grey dashed curve (\protect\gnuplotline{gray,dashed}). \textit{Bottom left.} Corresponding 3D stellar density profiles $\rho_\star(r)$.
\textit{Top right.} 2D stellar surface brightness profiles $\Sigma_\star(R)$. \textit{Bottom right.} Line-of-sight velocity dispersion profiles $\sigmalos(R)$. The luminosity-averaged line-of-sight velocity dispersion (Eq.~\ref{eq:sigmalos}) is virtually identical between the models and is shown as a grey band.}
\label{fig:IC_comp}
\end{figure*}

\subsection{Stellar Component Models}
\label{sec:cuspy_vs_cored_tracks}
Tidally stripped dwarf galaxies retain some memory of the properties of their initial stellar distribution \citepalias{ENIP2022}. We aim to explore  how different initial stellar distributions affect the tidal evolution of the remnant luminosity $L$, projected half-light radius $\Rh$ and line-of-sight velocity dispersion $\sigmalos$. In Sec.~\ref{subsubsec:exp2D}, \ref{subsubsec:exp3D} and \ref{subsubsec:log2Dstars}, we define the three different stellar density profiles that we adopt in our modelling. In Sec.~\ref{subsubsec:profile_comparison}, we briefly discuss the similarities and differences between these three models. Finally, in Sec.~\ref{subsec:stellar_profile_time_evolution}, we describe the time evolution of the surface brightness- and line-of-sight velocity dispersion profiles during tidal stripping. All stellar models discussed here provide a plausible description of the observed surface brightness profiles of faint Milky Way satellites (see Appendix~\ref{Appendix:ObservarionalConstraints} for a comparison against available observational data).

\subsubsection{2D Exponential -- \textup{\color{red}\texttt{exp2D}}}
\label{subsubsec:exp2D}
The surface brightness profiles of many Milky Way dwarf spheroidal galaxies are well approximated by 2D exponential profiles (see e.g. \citealt{Irwin1995} for fits to the Carina, Draco, Fornax, and Leo~I dwarf galaxies, or more recently \citealt{Wang2019} for Fornax, \citealt{ Sestito2023} for Ursa Minor and \citealt{Jensen2024} for multiple Milky Way satellites including Sculptor and several ultrafaint systems).
Motivated by this observational evidence, we adopt a 2D exponential as ``fiducial'' profile for our models, parameterized through
\begin{equation}
 \Sigma_\star(R)  = \Sigma_{0} ~ \exp\left( - R/ R_\star \right) 
 \label{eq:exp2D_Sigma}
\end{equation}
where $\Sigma_{0} =  {L} /  ({2\pi R_\star^2}) $, for a (2D) half-light radius of $\Rh \approx 1.68\,R_\star$. 
Under the assumption of spherical symmetry, we use the inverse Abel transform to reconstruct the corresponding 3D stellar density profile \citep[see, e.g.][Appendix 1B(4)]{BT87}, 
\begin{equation}
 \rho_\star(r) = - \frac{1}{\pi} \int_r^\infty  ~ \frac{{\diff \Sigma_\star(R)}/{\diff R}}{\sqrt{R^2 - r^2}} ~\diff R = \rho_\mathrm{ s\star} ~  K_0(r/R_\star)~,
 \label{eq:exp2D_rho}
 \end{equation}
where $K_0$ denotes the (zeroth-order) modified Bessel function of second type, and $\rho_\mathrm{s\star} =  \Sigma_0 / (\pi R_\star)$, for a (3D) half-light radius of  $\rh \approx 2.22\,R_\star$. Asymptotically, for $r\rightarrow0$ (see \citealt{AbramowitzStegun1972} Eq.~9.6.8),
\begin{equation}
 \rho_\star(r) ~\sim~ \rho_\mathrm{s\star} ~ \ln(R_\star/r)~,
 \label{eq:exp2D_asy}
\end{equation}
i.e., the 3D density distribution that generates a 2D exponential surface brightness profile diverges logarithmically at the centre.

\subsubsection{3D Exponential -- \textup{\color{tan1}\texttt{exp3D}}}
\label{subsubsec:exp3D}
We furthermore consider a stellar tracer that follows a 3D exponential density profile,
\begin{equation}
 \rho_\star(r) =   \rho_0 ~ \exp(-r/r_\star) 
 \label{eq:exp3D_rho}
\end{equation}
where $\rho_0 = {L} / (8\pi r_\star^3) $, for a 3D half-light radius of $\rh \approx 2.67\,r_\star$. The 2D projection of this profile gives
\begin{eqnarray}
 \Sigma_\star(R) &=& 2 \int_R^\infty  ~ \frac{\rho_\star(r)~r}{\sqrt{r^2 - R^2}} ~\diff r  =\Sigma_0  ~\frac{R}{r_\star} ~ K_1(R/r_\star)
 \label{eq:exp3D_Sigma}
\end{eqnarray}
where $K_1$ denotes the (first-order) modified Bessel function of second type, and $\Sigma_0 =  2 \rho_0 r_\star$, for a 2D half-light radius of $\Rh \approx 2.03\,r_\star$ (see \citealt{AbramowitzStegun1972} Eq.~9.6.9).

\subsubsection{2D Logarithmic Cusp -- \textup{\color{purple}\texttt{log2D}}}
\label{subsubsec:log2Dstars}
Finally, we consider a stellar tracer which, in 3D, diverges at the centre like an NFW dark matter profile, $\rho_\star(r) \sim r^{-1}$ for $r \ll r_\star$. For this model, we choose a stellar density profile formally identical to Eq.~\ref{eq:rho_asy},
\begin{equation}
 \rho_\star(r) = \rho_\mathrm{s\star} ~ \frac{r_\star}{r} ~ \exp\left( - r/r_\star \right) 
 \label{eq:log2D_rho}
\end{equation}
where $\rho_\mathrm{s\star} = {L} / (4\pi r_\star^3)$, with a 3D half-light radius of $\rh \approx 1.68\,r_\star$.
The 2D projection gives
\begin{equation}
 \Sigma_\star(R) = \Sigma_\mathrm{s} ~K_0(R/r_\star) ~,
 \label{eq:log2D_Sigma}
\end{equation}
where $K_0$ denotes the (zeroth-order) modified Bessel function of second type, and $\Sigma_\mathrm{s} = 2 \rho_\mathrm{s\star}r_\star$, for a 2D half-light radius of $\Rh \approx 1.26\, r_\star$. For $R\rightarrow0$,
\begin{equation}
 \Sigma_\star(R)  \sim \Sigma_\mathrm{s} \ln(r_\star/R)~,
\end{equation}
i.e. the (2D) surface brightness profile of an exponentially truncated \emph{cuspy} stellar (3D) density profile logarithmically diverges at the centre (see \citealt{AbramowitzStegun1972} Eq.~9.6.8).

\begin{figure*}
\centering
\hspace*{-1cm}\includegraphics[width=14.2cm]{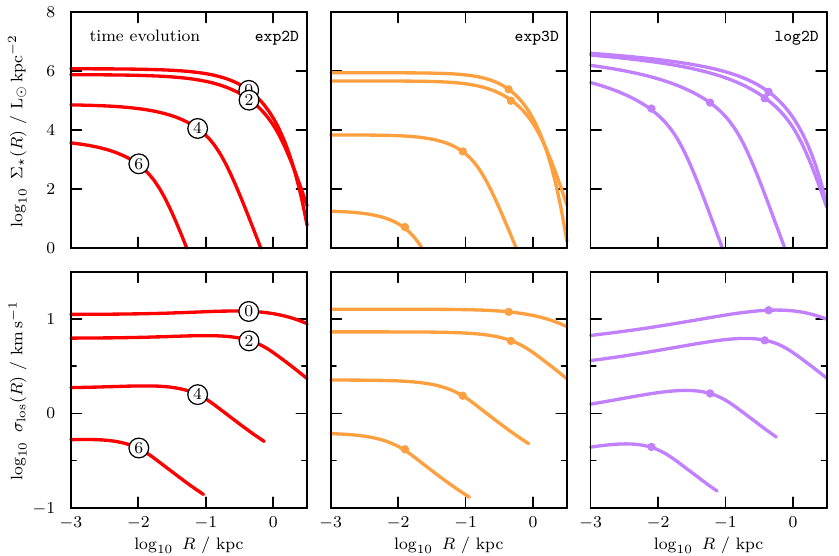}
\caption{\textit{Top panels.} Time evolution of the surface brightness profiles of the \texttt{exp2D} (\protect\gnuplotline{red}), \texttt{exp3D} (\protect\gnuplotline{tan1}) and \texttt{log2D} (\protect\gnuplotline{purple}) stellar models, with initial conditions as shown in Fig.~\ref{fig:IC_comp}. The \texttt{exp2D} tracer is identical to the model of Fig.~\ref{fig:2D_snapshots}, with selected snapshots, labelled $\protect\textcircled{0},\,...\,,\protect\textcircled{6}$, shown for fixed subhalo remnant masses of $\log_{10}\Mmx/M\mxzero=0,\,-2,\,-4\,,\,-6$. The label location coincides with the projected half-light radius. For the \texttt{exp3D} and \texttt{log2D} models, the half-light radius is marked using a filled circle. Note that the time evolution of the surface brightness profiles differs substantially between the three stellar models shown. \emph{Bottom panels}: Time evolution of the corresponding line-of-sight velocity dispersion profiles. Note that in the tidally limited regime, the velocity dispersion profile drops steeply beyond $\Rh$. }
\label{fig:corecusp_timeevo}
\end{figure*}

\subsubsection{Comparison of Initial Profiles}
\label{subsubsec:profile_comparison}
Here, we discuss the observable differences between the \texttt{exp2D}, \texttt{exp3D} and \texttt{log2D} stellar models defined in the previous three subsections. For reference, Table~\ref{tab:stellar_asymptotics} summarizes the central asymptotic behaviour of the three stellar models. 

In projection, the \texttt{exp2D}, \texttt{exp3D} models are \textit{cored}, i.e., $\diff \ln \Sigma_\star / \diff \ln R \rightarrow 0$ for $R\rightarrow 0$. In contrast, the surface brightness profile of the \texttt{log2D} model has a very shallow, logarithmically diverging central \textit{cusp}. The surface brightness profiles are shown in the top-right panel of Fig.~\ref{fig:IC_comp}, scaled to a half-light radius of $\Rh = 440\,\pc$. 

The \texttt{exp2D} model, shown in red, is identical to the example stellar tracer shown previously in Fig.~\ref{fig:stellar_dnde} and \ref{fig:2D_snapshots}. (Deprojected) 3D stellar density profiles are shown in the bottom-left panel of Fig.~\ref{fig:IC_comp}.

\begin{table}[t]
\caption{Central asymptotics for the 3D stellar density profiles $\rho_\star(r)$ and the 2D surface brightness profiles $\Sigma_\star(R)$ corresponding to the \texttt{exp2D}, \texttt{exp3D} and \texttt{log2D} models (for definitions, see Sec.~\ref{subsubsec:exp2D}, \ref{subsubsec:exp3D} and \ref{subsubsec:log2Dstars}, respectively). }
\label{tab:stellar_asymptotics}
\begin{center}
\begin{tabular}{l  @{\hspace{0.5cm}} l  @{\hspace{0.5cm}} l  @{\hspace{0.5cm}} l  }
\toprule
 \multirow{2}{*}{Model}        & \multirow{2}{*}{Eq.}                                                 & 3D Asymptotics                                  & 2D Asymptotics   \\
                               &                                                                      &    $r \ll \rh$                          &   $R \ll \Rh$ \\ \midrule
{\color{red}\texttt{exp2D}}    & \ref{eq:exp2D_Sigma}, \ref{eq:exp2D_rho}                             &   $\rho_\star(r) \sim \ln(R_\star/r)$   &   $\Sigma_\star(R) \rightarrow$ const  \\
{\color{tan1}\texttt{exp3D}}   & \ref{eq:exp3D_rho}, \ref{eq:exp3D_Sigma}                             &   $\rho_\star(r) \rightarrow$ const &   $\Sigma_\star(R) \rightarrow$ const  \\
{\color{purple}\texttt{log2D}} & \ref{eq:log2D_rho}, \ref{eq:log2D_Sigma}                             &   $\rho_\star(r) \sim 1/r$          &   $\Sigma_\star(R) \sim \ln(r_\star/R)$   \\ \bottomrule  
\end{tabular}
\end{center}
\end{table}

In 3D, only the \texttt{exp3D} model has a centrally finite density: the \texttt{exp2D} diverges logarithmically, and the \texttt{log2D} diverges with a power-law slope of $\diff \ln \rho_\star / \diff \ln r \rightarrow -1$  for $r\rightarrow 0$, like an NFW profile. Consequently, the stellar energy distribution of the \texttt{log2D} model has the same inner asymptotics as the dark matter when embedded in an NFW halo. This is shown in the top-left panel of Fig~\ref{fig:IC_comp} for a stellar tracer deeply embedded in an NFW halo ($\Rh/\rmx = 1/16$). Note that the energy distributions of the \texttt{exp2D} and \texttt{exp3D} models drop faster towards the most-bound state than the \texttt{log2D} model. 

Finally, in the bottom-right panel of Fig~\ref{fig:IC_comp} we show the line-of-sight velocity dispersion profiles of the three different stellar models, embedded in an NFW (subhalo) potential with parameters as in Table~\ref{tab:parameter_overview}. The luminosity-averaged line-of-sight velocity dispersion,
\begin{eqnarray}
 \langle \sigmalos^2 \rangle &=& \frac{2 \pi}{L} \int_0^\infty \Sigma_\star(R) \sigmalos^2(R) R~ \diff R  \label{eq:sigmalos} \\
                             &=& \frac{4\pi G}{3} \int_0^\infty r \nu_\star(r) M(<r)~\diff r  \label{eq:virial_disp} ~, \label{equ:sigmalos_virial}
\end{eqnarray}
shown as a grey-shaded band, is nearly identical for the three models. We denote by $\nu_\star(r)$ the (3D) stellar tracer density, normalized so that $4 \pi \int_0^\infty r^2 \nu_\star(r)\, \diff r = 1$, and $L = 2 \pi \int_0^\infty R \Sigma_\star \diff R$ is the total luminosity. $M(<r)$ is the total mass enclosed within  radius $r$. 

Eq.~\ref{eq:virial_disp} follows from the projected virial theorem \citep[see, e.g.][]{Amorisco2012, EPW18}, and guarantees that $\langle \sigmalos^2 \rangle$ is independent of anisotropies in the velocity dispersion. Note that a stochastic realization of the luminosity-averaged $\langle \sigmalos^2 \rangle$ is the only kinematic observable accessible for faint stellar systems that contain only a few stars sufficiently bright to accurately measure their radial velocities, and that this estimate may be artificially inflated if some of these stars are in binary systems. Distinguishing between these three models on kinematic grounds hence poses a considerable observational challenge.

\subsection{Tidal Evolution of the Stellar Components}
\label{subsec:stellar_profile_time_evolution}
As tides strip a dwarf galaxy, its surface brightness and velocity dispersion profiles evolve. For a given subhalo and orbit, the evolution depends on the shape of the initial stellar density profile, and on how deeply embedded the stellar profile is within its surrounding dark matter subhalo.

The top panel of Figure~\ref{fig:corecusp_timeevo} shows the tidal evolution of the surface brightness profiles for stellar tracers with initial conditions as in Fig.~\ref{fig:IC_comp}. Profiles are shown for subhalo remnant masses of $\log_{10}\Mmx/M\mxzero=0,\,-2,\,-4\,,\,-6$, as labelled in the left-hand panel. The location of the (2D) half-light radius is indicated by the label location in the left-hand panel, and by filled circles in the central and right-hand panel. 

For all three models shown the surface brightness profile is hardly affected by tides in the early stages of tidal evolution. Once the underlying subhalo has lost $\sim 99$ per cent of its initial characteristic mass, the tidal energy truncation reaches the stellar component (snapshot \textcircled{2}, compare with Fig.~\ref{fig:stellar_dnde}). From that point on, the surface brightness of the stellar component decreases with each further tidal energy truncation, and the half-light radius of the stellar component decreases. 

For the \texttt{exp2D} stellar model (left-hand panel), this evolution is illustrated also in Fig.~\ref{fig:2D_snapshots}. The surface brightness of the stellar component is shown in colour, whereas the dark matter surface density is shown in grey. After a slight initial expansion of the stellar half-light radius $\Rh$ (shown as a red circle), once the tidal energy truncation has reached the stellar component, dark matter and stars are truncated approximately at the same radius, and  the stellar half-light radius $\Rh$ evolves thereafter \textit{in sync} \citep{Kravtsov2010} with the characteristic size $\rmx$ (black circle) of the underlying dark matter subhalo. 

Returning to Fig.~\ref{fig:corecusp_timeevo}, we see that the surface brightness of the \texttt{exp2D} (red) and \texttt{exp3D} (orange) models drops much more rapidly during tidal stripping than the surface brightness of the \texttt{log2D} (purple) model. This disparate evolution is easily understood when looking at the underlying stellar energy distributions (Fig.~\ref{fig:IC_comp}). 

As tides truncate the dwarf galaxy in energy, the \texttt{log2D} model loses stars and dark matter at a similar pace, as its stellar energy distribution follows that of the dark matter towards the most-bound energy states. On the other hand, the most-bound energy states are hardly populated in the \texttt{exp3D} stellar model: its stellar energy distribution drops much more rapidly towards the most-bound states than that of the dark matter, making the stellar component vulnerable to full tidal disruption. The prospect of observing micro galaxies hence depends critically on how stars are distributed energetically within the dark matter subhalo. 

The bottom panel of Fig.~\ref{fig:corecusp_timeevo} shows the time evolution of the corresponding line-of-sight velocity dispersion profiles. The average velocity dispersion of all three stellar models drops monotonously with tidal mass loss: the evolution of the line-of-sight velocity dispersion depends only weakly on the shape of the initial stellar profile.

\section{Application to Local Group Satellites}
\label{sec:application_to_data}
We are now ready to apply the model outlined in Section~\ref{sec:tidal_evo_model} to predict the luminosity, structure, and kinematics of heavily stripped \textit{micro galaxies}, and to compare predicted properties with those of observed stellar systems in the Local Group. 

In particular, we aim to address the nature of recently-discovered stellar systems with half-light radii $1\lesssim \Rh/\pc \lesssim 30$ and luminosities of $10 \lesssim L/\Lsol \lesssim 10^3$ \citep{Torrealba2019clusters, Mau2020, Cerny22, Cerny23, Smith2023}. These systems populate a parameter space at the boundary between the globular cluster and dwarf galaxy regimes. If gravitationally bound, then, in principle, they could be one of the following three possibilities: 
\begin{itemize}[leftmargin=*,itemsep=0pt]
 \item[i)] self-gravitating star clusters devoid of dark matter, like globular clusters,
 \item[ii)] dark matter-dominated galaxies so deeply embedded within their dark matter haloes that their sizes and luminosities have not been affected by tides since formation, or
 \item[iii)] dark matter-dominated micro galaxies of tidal origin, formed through stripping of larger and more luminous progenitors.
\end{itemize}
We will in the following discuss these potential formation scenarios guided by the tidal evolutionary tracks computed for the \texttt{exp2D}, \texttt{exp3D} and \texttt{log2D} stellar models.

\subsection{Half-light Radii and Luminosities}
\label{sec:Rh_vs_L}

\begin{figure*}
 \gridline{\fig{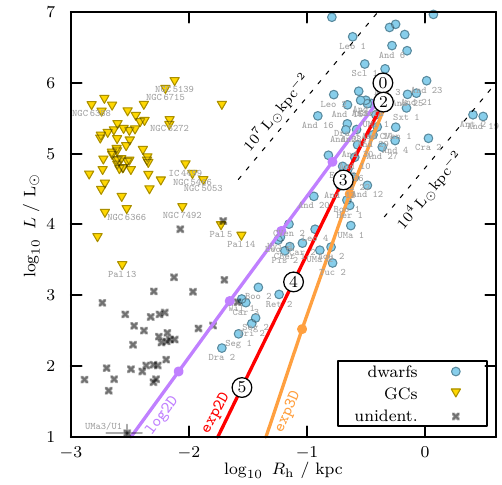}{8.5cm}{{(a)} Half-light radius $\Rh$ and luminosity $L$}
           \fig{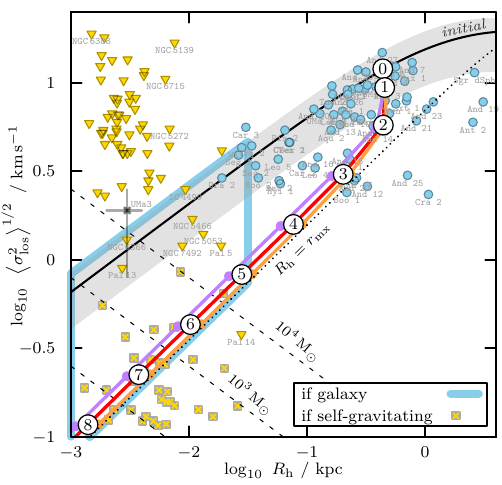}{8.5cm}{{(b)} Half-light radius $\Rh$ and velocity dispersion $\langle \sigmalos^2\rangle^{1/2}$ (Eq.~\ref{eq:sigmalos}) }
          }
 \gridline{\fig{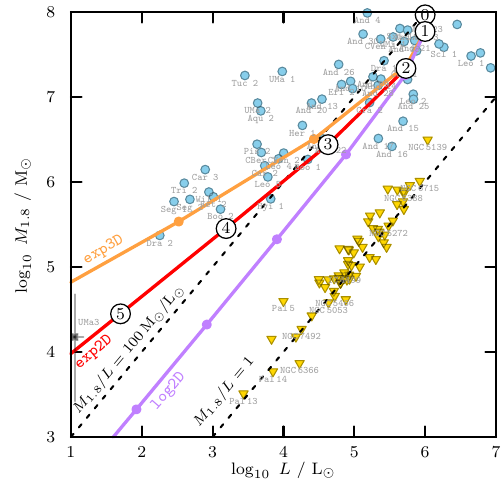}{8.5cm}{{(c)} Luminosity $L$ and dynamical mass $M_{1.8}=M(<1.8\,\Rh)$ (Eq.~\ref{eq:EPW18})}
           \fig{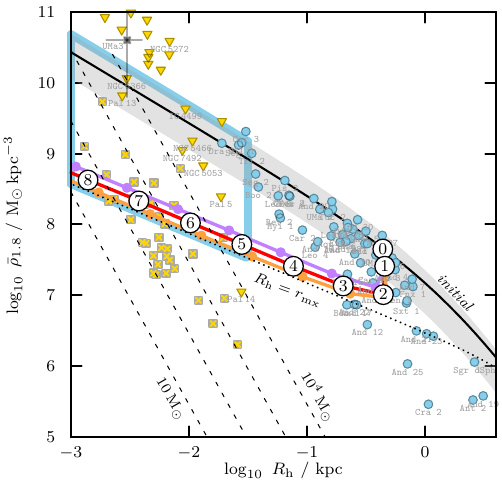}{8.5cm}{{(d)} Half-light radius $\Rh$ and mean density $\bar \rho_{1.8}=\bar \rho(<1.8\,\Rh)$ (Eq.~\ref{eq:density18})}
          }
\caption{
 \textbf{(a)} Luminosities $L$ and half-light radii $\Rh$ of Local Group dwarf galaxies~(\protect\gnuplotcircle[fill=dwarfcolor,draw=gray]) and globular clusters~(\protect\gnuplottriangle[fill=GCcolor,draw=gray]). Ambiguous objects at the boundary between the dwarf- and globular cluster regime are marked using grey crosses (\protect\gnuplotcross[draw=gray]). For references, see footnote~\ref{footnote:dsph_GC_data}. 
 Solid curves show tidal evolutionary tracks for the \texttt{exp2D} (\protect\gnuplotline[line width=2pt]{red}), \texttt{exp3D} (\protect\gnuplotline[line width=2pt]{tan1}) and \texttt{log2D} (\protect\gnuplotline[line width=2pt]{purple}) stellar models (with initial conditions as in Fig.~\ref{fig:IC_comp}). Snapshots corresponding to the 
 \texttt{exp2D} model, labelled $\protect\textcircled{0},\,...\,,\protect\textcircled{5}$, are highlighted for fixed subhalo remnant masses of $\log_{10}\Mmx/M\mxzero=0,\,-1,\,...\,,\,-5$.
 \textbf{(b)} Line-of-sight velocity dispersion $\langle \sigmalos^2 \rangle^{1/2}$ (Eq.~\ref{eq:sigmalos}) and projected half-light radius $\Rh$. The top black curve labelled ``initial'' shows the line-of-sight velocity dispersion expected for a stellar tracer of half-light radius $\Rh$ embedded in a $\sim10^9\,\Msol$ dark matter halo. Yellow crosses~(\protect\gnuplotcrosstwocolors) show velocity dispersions computed under the assumption of self-gravity for the ambiguous objects shown as grey crosses in the top-left panel. If instead these objects are dark matter-supported galaxies with $1\lesssim \Rh/\pc \lesssim 30$, their dispersion should fall in the region highlighted in blue~(\protect\gnuplotarea{draw=skyblue, fill=skyblue}). The region is computed assuming a dwarf galaxy progenitor embedded in a subhalo with a halo mass close to to the $z=0$ hydrogen cooling limit. For earlier formation redshifts, the lower bound shifts upwards, see Appendix~\ref{Appendix:FormationRedshift}. 
 \textbf{(c)}~Luminosities $L$ and dynamical masses $M_\mathrm{1.8}$ (Eq.~\ref{eq:EPW18}). Diagonal lines mark constant dynamical mass-to-light ratios of $M_{1.8}/L=1\,\Msol/\Lsol$ and $100\,\Msol/\Lsol$. %
 \textbf{(d)}~Projected half-light radii $\Rh$ and mean densities $\bar \rho_{1.8}$ (Eq.~\ref{eq:density18}) enclosed within a spherical radius of $1.8\,\Rh$.%
 \label{fig:data_vs_model_overview}
  }
\end{figure*}

Figure~\ref{fig:data_vs_model_overview}a shows a compilation{\footnotemark} of projected half-light radii, $\Rh$, and luminosities, $L$, of Local Group dwarf galaxies (blue circles) and globular clusters (yellow triangles). On average, at equal luminosity, globular clusters are significantly more compact than dwarf galaxies. For example, at $L=10^6\,\Lsol$, globular clusters span a range in sizes of roughly $1\sim10\,\pc$, more than an order of magnitude more compact than dwarfs of the same luminosity, $0.1\sim1\,\kpc$.

\footnotetext[\thefootnote]{The dwarf galaxy properties are taken from \citet{McConnachie2012} (version January 2021, with updates for Antlia~2 \citep{Ji2021}, Bootes~2 \citep{Bruce2023}, Crater~2 \citep{Ji2021}, Draco~2 \citep{Martin2016_Dra2, Longeard2018}, Tucana \citep{Taibi2020}, Tucana~2 \citep{Chiti2021}, And~19 \citep{Collins2020} and And~21 \citep{Collins2021}). For globular clusters, the data shown is from \citet{Harris1996} (version December 2010, with updated half-light radii and velocity dispersions for Pal-5 from \citealt{Kuzma2015,Gieles2021}; for NGC 2419 from \citealt{Baumgardt2009}, and for Pal-14 from \citealt{Hiker2006,Jordi2009}). Objects marked as ``unidentified'' at the boundary of the globular cluster- and dwarf galaxy regimes are as compiled in \citet{Cerny22, Cerny23}. Data for Ursa Major~3/Unions~1 is from \citet{Smith2023}.\label{footnote:dsph_GC_data}}

This clear separation in size becomes ambiguous at lower luminosities. Ultrafaint dwarfs like Bootes~2, Carina~3 and Draco~2 have half-light radii of $ 20$ to $30\,\pc$, not too different from extended Milky Way globular clusters like Palomar~5 or Palomar~14 (albeit at lower luminosity). Recently discovered objects, whose nature is still unclear, are labelled ``unidentified'' and depicted as grey crosses in Fig.~\ref{fig:data_vs_model_overview}a. They populate a region of parameter space at the boundary between the globular cluster and dwarf galaxy regimes. As an example, we show error bars for the faint Milky Way satellite Ursa Major~3/Unions~1 (UMa3/U1 for short), with a half-light radius of only $(3\pm1)\,\pc$, a total stellar mass of $M_\star = 16\plus{6}\minus{5}\,\Msol$, corresponding to a luminosity of $M_\mathrm{V}\approx +2.2\plus{0.4}\minus{0.3}\,\mathrm{mag}$ \citep{Smith2023}.

To address whether these ambiguous objects could be the remnants of tidally stripped dwarf galaxies, we show the evolution of three dwarf galaxy models in Fig.~\ref{fig:data_vs_model_overview}a, computed using the method outlined in Sec.~\ref{sec:tidal_evo_model}.
As an example, we assume that the progenitor dwarf galaxy is initially embedded in a dark matter subhalo with structural properties as listed in Tab~\ref{tab:parameter_overview}, with a viral mass close to the redshift $z=0$ hydrogen cooling limit. The systematics arising from earlier formation redshifts and lower halo masses are studied in Appendix~\ref{Appendix:FormationRedshift}. The example progenitor considered here has an initial luminosity of $L = 10^6\,\Lsol$, a half-light radius of $R_\mathrm{h} = 440\,\pc$ and a velocity dispersion $\sigma_\mathrm{los} \approx 12\,\kms$, chosen to roughly resemble classical Milky Way satellites like Ursa Minor, Sextans or Sculptor.

The evolution of size and luminosity for a dwarf galaxy with an \texttt{exp2D} surface brightness profile is shown in red (identical to the model shown in Fig.~\ref{fig:stellar_dnde} and \ref{fig:2D_snapshots}, with circled numbers $\protect\textcircled{0},\,...\,,\protect\textcircled{6}$ corresponding to fixed subhalo remnant masses of $\log_{10}\Mmx/M\mxzero=0,\,-1,\,...\,,\,-6$). 
As tides remove stars from the \texttt{exp2D} model, its size and luminosity drop. The evolution moves the dwarf galaxy model roughly along the observed mass--size relation of dwarf galaxies.
The evolution of a dwarf galaxy with an initial half-light radius and luminosity different from the example model shown in Fig.~\ref{fig:data_vs_model_overview}a can be roughly estimated by shifting the tidal evolutionary track within the size-luminosity plane. 

As discussed in Section~\ref{sec:cuspy_vs_cored_tracks}, the evolution of size and luminosity crucially depends on the shape of the initial stellar density profile, which in turn depends on the degree to which stellar binding energies extend to the most-bound energy states in the subhalo. The evolution of the \texttt{exp3D} model is shown in orange. For this model, the luminosity drops faster with radius than in the case of the \texttt{exp3D} model. 

On the other hand, for the \texttt{log2D} model (shown in purple), the drop in luminosity is less steep\footnote{Some simple analytical insight in the disparate evolution of stellar models with \texttt{exp2D}, \texttt{exp3D} and \texttt{log2D} density profiles can be gained from the tidally limited regime, when the stellar component and the subhalo are trimmed down to similar sizes and evolve in sync ($\Rh \propto \rmx$ and $M_\mathrm{h} \propto \Mmx$). Calling $\alpha_\star$ the slope of the stellar energy distribution $\diff N_\star / \diff \E \propto \E^{\alpha_\star}$ for $\E \ll 1$, then $ L \propto M_\mathrm{h}^{(\alpha_\star+1)/2}$ (assuming NFW haloes, where $\diff M / \diff \E \propto \E$ for $\E \ll 1$). Combining this with the tidal track of Eq.~\ref{eq:Eqn21track} with $M_\mathrm{h} \propto \Rh^{2\beta+1}$ gives $ \diff \ln L / \diff \ln \Rh = (2\beta+1)(\alpha_\star+1)/2$. Hence, the evolution of luminosity and size of a stellar tracer component is directly related to how stars populate the most-bound energy states of the subhalo, parameterized through the log-slope $\diff \ln (\diff N / \diff \E) / \diff \ln \E = \alpha_\star$ for $\E \ll 1$.}: tidal evolution shifts the dwarf galaxy model roughly along the observed mass--size relation of dwarf galaxies, and, for the example initial conditions chosen here, the stripped model roughly approaches the region of parameter space occupied by the unidentified systems.

It is clear that the majority of the unidentified systems depicted as grey crosses in Fig.~\ref{fig:data_vs_model_overview}a cannot be reached by tidally stripping progenitor systems with initial luminosites and sizes comparable to those assumed in the example above. A progenitor whose tidal descendent is consistent with the unidentified systems can be found by shifting the tidal tracks upwards and to the left (Leo 1 could be one example). We may conclude that, if the unidentified systems are indeed micro galaxies, then their progenitors must have been systems of higher luminosity and/or surface brightness than the typical dSphs, and their stellar density must be closer to that of the \texttt{log2D} model than \texttt{exp3D}.

The tidal tracks shown in Fig.~\ref{fig:data_vs_model_overview}a  highlight the fact that tidal evolution in the size/luminosity plane is very sensitive to the underlying distribution of stellar binding energies. Further constraints are needed to either rule out or to confirm conclusively the possibility that many of the unidentified systems are indeed micro galaxies.

\subsection{Half-light Radii and Velocity Dispersions}
\label{subsec:Rhvssigma}

We now discuss to what extent measurements of the line-of-sight velocity dispersion $\langle \sigmalos^2 \rangle^{1/2}$ (Eq.~\ref{eq:sigmalos}) can help to constrain the nature of the unidentified stellar systems. Fig.~\ref{fig:data_vs_model_overview}b shows the half-light radii $\Rh$ and line-of-sight velocity dispersions $\langle \sigmalos^2 \rangle^{1/2}$ for the same dwarf galaxies and globular clusters as in Fig.~\ref{fig:data_vs_model_overview}a. Dwarf galaxies, on average, are more extended than globular clusters at equal velocity dispersion.

\subsubsection{Self-gravitating Star Clusters}
For most of the objects marked as ``unidentified'' in Fig.~\ref{fig:data_vs_model_overview}a, no robust measurement of velocity dispersion is available to date.
If these objects were self-gravitating and devoid of dark matter, their velocity dispersion could be estimated from their half-light radius $\Rh$ and their total stellar mass $M_\star$.
For a 3D exponential stellar density profile (for definition see Sec.~\ref{subsubsec:exp3D}) with $\Rh \approx 2.03\,r_\star$ and total stellar mass $M_\star$, Eq.~\ref{eq:virial_disp} gives
\begin{equation}
 \langle \sigmalos^2 \rangle = (5/96) ~G  M_\star / r_\star~. 
 \label{eq:sigmalos_exp}
\end{equation}
For each ``unidentified'' system shown in Fig.~\ref{fig:data_vs_model_overview}a, we use Eq.~\ref{eq:sigmalos_exp} to estimate the (virial) velocity dispersion the object would have in the absence of dark matter. For this calculation, we assume a mean stellar mass-to-light ratio of $\Upsilon \approx 1.6$ \citep{Woo2008}. The results of this exercise\footnote{The velocity dispersion estimates for self-gravitating clusters are fairly insensitive to the choice of the underlying density profile: for Plummer spheres, we find $ \langle \sigmalos^2 \rangle = (\pi/32) ~G  M_\star / \Rh $, where as before by $M_\star$ we denote the total stellar mass, and by $\Rh$ the projected half-light radius. For a given $M_\star$ and $\Rh$, the dispersion $\langle \sigmalos^2 \rangle^{1/2}$ differs by ${\sim}4$ per cent between a Plummer sphere and a 3D exponential profile.} are marked as yellow crosses in Fig.~\ref{fig:data_vs_model_overview}b. 
For reference, dashed diagonal lines in Fig.~\ref{fig:data_vs_model_overview}b are also computed from Eq.~\ref{eq:sigmalos_exp}, for constant stellar masses of $M_\star = \Upsilon L = 10^2$, $10^3$ and $10^4\,\Msol$.

\subsubsection{Dark Matter-dominated Objects}
For the case of dark matter-dominated dwarf galaxies, the gravitational potential is sourced primarily by the dark matter component. Their stellar velocity dispersions depend mainly on the mass of the halo, and the degree of spatial segregation between stars and dark matter.

In the LCDM cosmology, dwarf galaxies are expected to form within a relatively narrow range of halo masses \citep[see e.g.][]{Guo2010, Guo2011, Fattahi2018}. An estimate of the characteristic minimum halo mass may be obtained from the constraint that, to form stars, gas must be able cool in presence of the cosmic UV background. Only haloes with masses above the \emph{hydrogen cooling limit} provide the necessary conditions for the onset of star formation. 

At redshift $z=0$, this limiting virial mass is roughly $\sim 5\times10^9\,\Msol$ \citep[see, e.g.,][]{Benitez-Llambay2020}. The velocity dispersion of an exponential stellar profile embedded in an average-concentration NFW halo with a halo mass close to the $z=0$ hydrogen cooling limit ($r\mxzero=7\,\kpc$, $V\mxzero=35\,\kms$, see Table~\ref{tab:parameter_overview}) is shown as a black solid curve in Fig.~\ref{fig:data_vs_model_overview}b, labelled ``initial''. This curve is computed from Eq.~\ref{eq:virial_disp}, with the integral evaluated numerically. We discuss the redshift-dependence of these initial conditions in Appendix~\ref{Appendix:FormationRedshift}.

More generally, cosmological simulations suggest that satellites of Milky Way-like host galaxies formed within dark matter haloes of peak circular velocity $20\lesssim \Vmx/\kms \lesssim 40$ \citep[e.g.,][]{Fattahi2018}.
This range of potential initial haloes is indicated by the grey band in Fig.~\ref{fig:data_vs_model_overview}b, which includes $\pm0.15\,\dex$ variation in concentration around the mean.  Note that many dSphs of the Local Group fall within this band, which suggests that they have experienced only minor tidal perturbations. 

The present-day sizes and dispersions of dwarf galaxies and their remnants depend on their properties at formation, and on their tidal history. In particular, dwarf galaxies that experienced little tidal mass loss would be located within the grey band of Fig.~\ref{fig:data_vs_model_overview}b.

Tidal stripping reduces the stellar velocity dispersion, providing a potential explanation as to why some observed dwarfs are found below the grey band. 
To illustrate this, we show the tidal evolution of half-light radius $\Rh$ and velocity dispersion $\langle \sigmalos^2 \rangle^{1/2}$ of the three example dwarf galaxy models with \texttt{exp2D}, \texttt{exp3D} and \texttt{log2D} stellar tracers using red, orange and purple curves in Fig.~\ref{fig:data_vs_model_overview}b. 
Both models are initially located at the position marked \textcircled{0}, and are embedded in the same dark matter halo.
As previous, circled numbers $\protect\textcircled{0},\,...\,,\protect\textcircled{6}$ correspond to the snapshots shown in Fig.~\ref{fig:2D_snapshots} at fixed subhalo remnant masses of $\log_{10}\Mmx/M\mxzero=0,\,-1,\,...\,,\,-6$.

In the early stages of tidal evolution, the tidal energy truncation does not reach the binding energies of the stars (compare with Fig.~\ref{fig:stellar_dnde}). Dark matter on weakly-bound orbits is removed by tides. Some of these orbits do contribute to the density within the stellar radii: after their removal, the velocity dispersion of an embedded stellar tracer drops. This causes the initial near-vertical evolution in the size--velocity dispersion plane. 

Once the tidal energy truncation has reached the stellar component (in the example shown, this corresponds roughly to snapshot \textcircled{2}), size and velocity dispersion of stars and dark matter evolve in sync, both decreasing monotonously\footnote{For heavily stripped stellar systems in LCDM, a decrease in the line-of-sight velocity dispersion is always accompanied by a decrease in size of the bound component. The velocity dispersions and sizes of the ``feeble giant'' \citep{Torrealba2016} satellites Ant~2, Cra~2, And~19, And~25 can therefore not be reached through tides from initial conditions within the grey band of Fig.~\ref{fig:data_vs_model_overview}b (assuming that they are bound objects in dynamical equilibrium), see \citet{Borukhovetskaya2022}, \citetalias{ENIP2022}.}. 

This is the regime we refer to as \emph{tidally limited} in \citetalias{ENIP2022}. In this regime, for the \texttt{exp2D} stellar tracer, we find $\Rh/\rmx \approx 0.8$. Similarly, we find $\Rh/\rmx \approx 0.9$ for the \texttt{exp3D} stellar tracer, and $\Rh/\rmx \approx 0.6$ for the \texttt{log2D} model. For all three models, in the tidally limited regime\footnote{A simple analytical estimate for the line-of-sight velocity dispersion in the tidally limited regime can be obtained from Eq.~\ref{equ:sigmalos_virial}, assuming as stellar profile a 3D exponential sphere (Eq.~\ref{eq:exp3D_rho}, where $r_\star \approx \Rh/2.03$), and as underlying potential a truncated NFW cusp (Eq.~\ref{eq:rho_asy}, where $r_\mathrm{cut} \approx \rmx/1.79$, and $\sqrt{G M_\mathrm{cut} / r_\mathrm{cut}} \approx \Vmx/0.546$). For this choice of stellar profile and dark matter potential, the integral in Eq.~\ref{equ:sigmalos_virial} has an analytical solution, and
\begin{equation}
\langle \sigmalos^2 \rangle = \frac{G M_\mathrm{cut}}{6\,r_\mathrm{cut}}~ \frac{ (r_\star/r_\mathrm{cut}) \left(3 + r_\star/r_\mathrm{cut}\right)}{\left(1 + r_\star / r_\mathrm{cut}\right)^3} ~.
\label{eq:sigma_trunc}
\end{equation}
For a tidally limited galaxy, $\Rh \approx \rmx$, and Eq.~\ref{eq:sigma_trunc} yields $\langle \sigmalos^2 \rangle^{1/2} \approx  0.5\, \Vmx$.}, $\langle \sigmalos^2 \rangle^{1/2}/\Vmx \approx 0.5 $. 

The initial shape of the stellar density profile appears to play only a secondary role in the evolution of half-light radius and luminosity-averaged line-of-sight velocity dispersion; the model predictions for the \texttt{exp2D}, \texttt{exp3D} and \texttt{log2D} stellar tracers in Fig.~\ref{fig:data_vs_model_overview}b differ very little.

We can now return to the objects marked ``unidentified'' in Fig.~\ref{fig:data_vs_model_overview}a. If these objects were dark matter-dominated and only weakly affected by tides, they would have velocity dispersions that fall in the grey-shaded band of Fig.~\ref{fig:data_vs_model_overview}b. If instead these objects were subjected to strong tides, their velocity dispersions should fall within the region highlighted in blue in Fig.~\ref{fig:data_vs_model_overview}b. In either of the two cases, for many of the ``unidentified'' objects, the predicted velocity dispersions for the dark matter-dominated formation scenario are higher than those expected for self-gravitating objects devoid of dark matter (yellow crosses). As an example, for the case of the faint Milky Way satellite Ursa Major~3/Unions~1, in Fig.~\ref{fig:data_vs_model_overview}b, we show a velocity dispersion estimate of $\sigmalos=1.9\plus{1.4}\minus{1.1}$ (obtained after removing the furthest outlier from the sample of measured velocities as described in \citealt{Smith2023}). Note that this dispersion estimate may still be inflated by the presence of binary stars. Taking for now the available dispersion estimate at face value, it appears that UMa3/U1 is consistent with being a dark matter-dominated system. If in dynamical equilibrium and devoid of dark matter, UMa3/U1 would be expected to have a velocity dispersion of only ${\sim}50\,\mathrm{m}/\mathrm{s}$ \citep{ENSM24}, below the range plotted in Fig.~\ref{fig:data_vs_model_overview}b. 

These results suggest that accurate velocity dispersion estimates of micro galaxy candidates can help to distinguish them from globular cluster remnants: If devoid of dark matter, each ``ambiguous'' object should have a velocity dispersion that is roughly consistent with that expected for a self-gravitating object of its luminosity. On the other hand, if dark matter-dominated, the velocity dispersion should be in excess of that expected for a self-gravitating object of its luminosity. Crucially, our model predictions for the size--velocity dispersion relation of micro galaxies are relatively insensitive to the (unknown) initial energy distribution of stars within the dark matter halo. The required level of measurement accuracy is, however, challenging, with (``virial'', Eq.~\ref{eq:virial_disp}) velocity dispersions in the dark matter-dominated case ranging from several hundred $\mathrm{m}\mathrm{s}^{-1}$ to a few $\kms$. Accurate multi-epoch spectroscopy will be required to overcome uncertainties due to the potential contribution of binaries to the observed dispersions \citep[see, e.g.,][]{McConnachieCote2010, Koposov2011, Minor2019}.

\subsection{Dynamical Mass-to-light Ratio}
As discussed in the previous section, micro galaxies would differ from globular clusters in their dark matter content.
The virial theorem can be used to estimate the dynamical mass enclosed within the luminous radius of a stellar system through the combined measurement of half-light radius $\Rh$ and luminosity-averaged velocity dispersion $\langle \sigmalos^2 \rangle$ \citep{Illingworth1976,Merritt1987,Amorisco2012}.
We will in the following adopt the mass estimator derived in \citet{EPW18}, which is insensitive to anisotropies in the velocity dispersion, and minimizes uncertainties introduced by the (unknown) dark matter density profile shape and extent. 
The estimator yields the mass enclosed within a spherical radius of $\sim1.8\,\Rh$, 
\begin{equation}
 M_{1.8} \equiv M(<1.8\,\Rh) \approx 3.5 \times 1.8\,\Rh~\sigmalos^2~ G^{-1}  ~.
 \label{eq:EPW18}
\end{equation}
In Fig.~\ref{fig:data_vs_model_overview}c, we compare the dynamical mass $M_{1.8}$ of Local Group dwarf galaxies and globular clusters against their total luminosity $L$. 
The globular cluster luminosity and dynamical mass are distributed with little scatter around a line of constant dynamical mass-to-light ratio $M_{1.8}/L \approx 1\,\Msol/\Lsol$. Dwarf galaxies, on the other hand, span a wide range of dynamical mass-to-light ratios, $10 \lesssim (M_{1.8}/L)(\Msol/\Lsol)^{-1} \lesssim 10^4$.
For the case of the faint stellar system UMa3/U1, taking the velocity dispersion estimate of $\sigmalos \approx 1.9\,\kms$ at face value, we find a dynamical mass-to-light ratio of order ${\sim}10^3$, suggesting that UMa3/U1 is consistent with being heavily dark matter-dominated.

The tidal evolution of the dynamical mass-to-light ratio depends on the shape of the stellar density profile. The dynamical mass-to-light ratio of the \texttt{exp2D} and \texttt{exp3D} models (shown in Fig.~\ref{fig:data_vs_model_overview}c as red and orange curves, respective) gradually increases in the regime of heavy mass loss, as the stellar energy distribution of these two models drops more steeply towards the most-bound states than that of the surrounding dark matter halo (see Fig.~\ref{fig:IC_comp}). As the tidal mass loss progresses, 
the \texttt{exp2D} and \texttt{exp3D} models become more and more dark matter dominated. 
In contrast, the mass-to-light ratio of the \texttt{log2D} stellar model (with a stellar energy distribution which traces that of the dark matter) converges during tidal stripping to a constant value. This asymptotic mass-to-light ratio is lower than the mass-to-light ratio at accretion. Yet, for the examples considered here, the dwarf galaxies do remain dark matter dominated throughout their evolution. Micro galaxies can therefore be distinguished from globular clusters by their dark matter content, either directly, through measurements of their stellar velocity dispersions (see Sec.~\ref{subsec:Rhvssigma}), or indirectly, by studying their susceptibility to tidal forces along their orbit (see Sec.~\ref{Sec:meanDensity}).

\subsection{Enclosed Mean Density}
\label{Sec:meanDensity}
Using the enclosed dynamical mass $M_{1.8}$ defined by Eq.~\ref{eq:EPW18}, we can estimate the density of a stellar system averaged within a spherical radius of $1.8\,\Rh$,
\begin{equation}
 \bar \rho_{1.8} \equiv M_{1.8} ~ (1.8\,\Rh)^{-3} ~  (4\pi/3)^{-1}~. 
 \label{eq:density18}
\end{equation}
Fig.~\ref{fig:data_vs_model_overview}d shows the enclosed mean densities $\bar \rho_{1.8}$ of Milky Way satellites as well as their half-light radii $\Rh$. 
All elements of Fig.~\ref{fig:data_vs_model_overview}d are analogous to Fig.~\ref{fig:data_vs_model_overview}b, but expressed in terms of density. 
This allows for an intuitive description of the tidal evolution of dwarf galaxies. In the early stages of tidal evolution, luminosity and extent of the stellar component are hardly affected. Yet, the mean density enclosed within the luminous radii drops, as dark matter on weakly-bound orbits is stripped (snapshot $\textcircled{1}$). Once the tidal energy truncation affects the stellar component, its size decreases (snapshot $\textcircled{2}$). At the same time, its mean density gradually increases: the smaller the remaining stellar system, the higher the average density of the underlying cuspy dark matter halo enclosed within the luminous radii.

Returning to the objects marked as ``unidentified'' in Fig.~\ref{fig:data_vs_model_overview}a, we note that, for many of these objects, in the absence of dark matter their mean densities (yellow crosses) would be substantially lower than the model prediction for dark matter-dominated micro galaxies (blue region). For the case of the faint satellite UMa3/U1, the mean density estimated for a velocity dispersion of $\sigmalos \approx 1.9\,\kms$ is roughly consistent with the LCDM prediction. If devoid of dark matter, the low stellar mass of UMa3/U1 ($M_\star \approx 16\,\Msol$) would result in a mean density well below the LCDM prediction, falling roughly on the lowermost dashed curve in Fig.~\ref{fig:data_vs_model_overview}d. 

Remarkably, if self-gravitating, the mean densities of several of the ``unidentified'' systems are comparable to the mean density of the Milky Way at the solar circle, $\bar \rho_\mathrm{MW}(R_\odot) \approx 5 \times 10^7\,\Msol\,\kpc^{-3}$ (assuming $R_\odot=8.3\,\kpc$ and a circular velocity of $240\,\kms$). This in turn implies that these objects are likely vulnerable to Galactic tides, which offers an alternative probe to constrain their potential dark matter content.

Studying the vulnerability to tidal disruption of stellar systems is of particular relevance for those objects where accurate estimates of the internal velocity dispersion are unavailable, or deemed unreliable because of the potential contribution of binary stars to the measured dispersion. 
As long as the Galactocentric velocity of a stellar system can be measured and its orbit constrained, the strength of the tidal field experienced on that orbit can serve to inform whether the presence of dark matter is necessary to avoid full tidal disruption over a time span comparable to the stellar ages in the system. We explore this method with more detail in \citet{ENSM24} for the example of the Milky Way satellite Ursa Major~3/Unions~1 \citep{Smith2023}, concluding that only if stabilized by the presence of dark matter could the system survive for more than two radial orbital periods on its current orbit.

\subsection{How Many?}
The model outlined in Sec.~\ref{sec:tidal_evo_model} enables predictions of the luminosities, sizes and velocity dispersions of heavily stripped dwarf galaxies -- but not their expected abundance in the Milky Way. Simulation studies \citep[e.g.][]{vdBTormen2005,Springel2008,Libeskind2010,EPLG17} and semi-analytical frameworks \citep[e.g.][]{Han2016,Green2021_mass_function} have been used to estimate the subhalo mass function, and/or the radial distribution of subhaloes within a larger host. 
\citet{Tolerud2008}, e.g., suggest that the Milky Way could host between ${\sim}300$ and ${\sim}600$ luminous satellites within $400\,\kpc$ of the galactic centre. Similarly, \citet{Manwadkar2022} compute a total of roughly $\sim440$ satellites with $M_V<0$ (i.e. $\log_{10} L/\Lsol \gtrsim 1.9$) and half-light radii $\Rh \gtrsim 10\,\pc$ within $300\,\kpc$. 
\citet{Ahvazi2023} estimate a total of $\sim300$ satellites with $M_V<0$ within $300\,\kpc$, while \citet{Newton2018} argue for a somewhat lower value of ${\sim}120$ satellites. The actual number of heavily stripped satellites in the inner regions of the Milky Way is shown to be sensitive to its detailed mass assembly history \citep[see e.g.][]{Bose2020}.

The number of luminous satellites is determined by the interplay of the subhalo mass function, and the lowest halo mass that allows star formation. For the Aquarius~A halo, \citet{Springel2008} measure $\diff \ln N_\mathrm{sub}/ \diff \ln M_\mathrm{sub} \approx -1.9$, i.e., roughly speaking, the number of subhaloes per log-spaced mass bin is inversely proportional to the subhalo mass. Only haloes above some threshold mass allow hydrogen gas to cool in presence of the cosmic UV background, and in turn, to collapse and form stars \citep[see, e.g.][]{Bullock2000, Gnedin2000}. The critical mass model discussed in \citet{Benitez-Llambay2020}, for example, implies a redshift-dependent minimum halo mass for star formation which, at $z=0$, equals $\sim5 \times 10^9\,\Msol$ (see Appendix~\ref{Appendix:FormationRedshift} for a discussion on the redshift dependence of our results). Galaxies which formed their stars prior to reionization may have formed in haloes of substantially smaller masses: e.g., \citet{BovillRicotti2009} suggest a threshold of $\sim 2 \times 10^8\,\Msol$, while \citet{Manwadkar2022} suggest that half of the haloes with a peak mass of only ${\sim}4\times10^7\,\Msol$ may host a galaxy. 

For a numerical example, we use as before the Aquarius-A2 main halo as a guide (see Sec.~\ref{sec:arti_disrupt}). Within the $z=0$ virial radius ($r_{200}\approx 246\,\kpc$), we count a total of $N_\mathrm{sub} = 16$, $65$, $152$ and $347$ subhaloes with peak masses above $5\times10^9\,\Msol$, $1\times10^9\,\Msol$, $5\times10^8\,\Msol$ and $2\times10^8\,\Msol$, respectively. To roughly estimate the number of subhaloes that reach the inner regions of a Milky Way-like dark matter halo, we return to study the orbits of subhaloes in the Aquarius-A2 merger tree. Using the same setup as in Sec.~\ref{sec:arti_disrupt}, we treat subhaloes as point-masses, and integrate their orbits from their respective redshift of accretion $z_\mathrm{acc}$ till $z=0$. We assume a time-evolving, analytical potential fitted to the A2 main halo (see Sec.~\ref{subsec:raporperi} for details). This allows us to follow the orbits of all subhaloes resolved at accretion till $z=0$, without losing subhaloes to artificial disruption. We find that within a Galactocentric radius of $r_{200}\approx 246\,\kpc$ there are $31$ subhaloes with peak virial masses $M_\mathrm{peak} \ge 1\times 10^9 \,\Msol$ on orbits with pericentric distances $\rperi \le 20\,\kpc$, and $16$ subhaloes have $\rperi \le 10\,\kpc$. Decreasing the mass threshold to $M_\mathrm{peak} \ge 2 \times 10^8 \,\Msol$ increases the number of subhaloes with $\rperi \le 20\,\kpc$ to 155, and the number of those with $\rperi \le 10\,\kpc$ to 82. Note that cosmological variance will introduce a substantial halo-to-halo scatter of these numbers, in particular at the higher-mass end (see figure~8 in \citealt{Springel2008} comparing the subhalo abundance in the Aquarius~A--F main haloes).

The strong dependence of the subhalo abundance on the choice of threshold mass is a direct consequence of the steep underlying subhalo mass function. Observational constraints on the abundance of micro galaxies would inform us equally about the number of dark matter subhaloes in the inner Milky Way, and about the minimum mass of star-forming subhaloes.

\section{Summary and Conclusions}
\label{sec:conclusions}
In the LCDM cosmology, dark matter haloes are predicted to have remarkably dense centres where the density profiles formally diverge as $\diff \ln \rho / \diff \ln r = -1$ for $r \rightarrow  0$. These density cusps render LCDM subhaloes resilient to the effect of tides, and prevent the full tidal disruption of subhaloes and embedded dwarf galaxies. The hierarchical accretion history of the Milky Way may therefore give rise to a population of heavily stripped ``micro galaxies'', i.e., co-moving groups of stars, held together and shielded from Galactic tides by a surrounding dark matter subhalo. 

The resilience to full tidal disruption distinguishes LCDM from other dark matter theories, many of which predict haloes with central constant-density cores, and/or lower mean central densities, such as ultralight-particle DM models ("fuzzy dark matter", FDM), or self-interacting dark matter (SIDM) models (prior to the onset of core-collapse).

In this study, we have used an empirical framework to model the evolution of size, luminosity and velocity dispersion of stellar systems embedded in LCDM subhaloes, following their tidal evolution over many orders of magnitude in mass, luminosity, and size. We discuss our findings in the context of recently-discovered faint stellar systems with structural properties at the boundary of the globular cluster- and dwarf galaxy regimes.

\subsection{Caveats}
We make several modelling assumptions in this study, which we summarize here as caveats to our conclusions. 
\begin{itemize}[leftmargin=*, itemsep=0pt]
 \item[i)] We assume that dark matter haloes prior to accretion are well-approximated by NFW profiles \citep{Navarro1996a, Navarro1997}, which centrally diverge as $\rho(r)\sim r^{-1}$. High-resolution cosmological simulations suggest that Einasto profiles, with a centrally-decreasing power-law slope, provide a slightly better description of the inner regions of CDM haloes \citep{NavarroLudlow2010, LudlowNavarro2013, Wang2020}, whereas ``prompt cusps'' \citep{Delos2023gamma} are argued to cause a steep $\rho(r)\sim r^{-3/2}$ divergence in the innermost regions  \citep{Ishiyama2010, OgiyaHahn2018, DelosWhite2023}. 
 \item[ii)] Baryonic feedback processes may erode dark matter cusps, rendering CDM subhaloes more vulnerable to tidal mass loss and disruption. The existence of ``micro galaxies'' hinges on the dark matter \emph{cusps} being intact, and stars populating the most-bound energy states within them. 
 \item[iii)] We model stars as tracers of the underlying dark matter potential. Stellar systems that are gravitationally dominant over a dark matter component, or fully self-gravitating, cannot be described by our framework. 
 \item[iv)] The evolution of both dark matter and stars assumes spherical, dispersion-supported systems with isotropic kinematics. The tidal evolution of rotating or strongly anisotropic systems cannot be modelled with the current framework. 
 \item[v)] We base our analysis on tidal evolutionary tracks derived from high-resolution $N$-body simulations, assuming a power-law extrapolation  for remnant masses that are unresolved in the simulations. Recent (semi-)analytical models propose tidal tracks with (slightly) different asymptotics \citep{Amorisco2021,Stuecker2023}.
 \item[vi)] The concentration--mass--redshift relation underlying this study is calibrated to dark matter-only cosmological simulations, ignoring the effect of baryonic feedback on halo concentrations. Note however that for the halo masses of interest in this study, comparison of the scatter in halo concentration in our initial conditions against hydrodynamical simulations shows good agreement, see Appendix~\ref{appendix:scatter}.
 \item[vii)] Finally, we assume a redshift-dependent minimum halo mass for star formation following the \citet{Benitez-Llambay2020} model (after reionization) and \citet{Tegmark1997} model (prior to reionization). We discuss the systematics arising from lowering the threshold halo mass for star formation in Appendix~\ref{Appendix:FormationRedshift}.
\end{itemize}

\pagebreak
\subsection{Conclusions}
Our main conclusions are summarized below. 
\begin{itemize}[leftmargin=*, itemsep=0pt]
 \item[i)] Current cosmological simulations cannot accurately predict the structure, kinematicsm and abundance of subhaloes and embedded dwarf galaxies in the innermost regions of the Galaxy. Insufficient $N$-body particle number and spatial resolution drive the artificial disruption of substructures in the inner regions of simulated Milky Way--like haloes. A route to circumvent such resolution limits are analytical models like the one discussed in this work. 
 \item[ii)] Consistent with earlier work, we show that, if a stellar tracer populates the energy states of a CDM halo all the way down to the most-bound state, then smooth tidal fields cannot fully disrupt that stellar tracer. This suggests the possible existence of micro galaxies, co-moving stellar systems of low total luminosity, stabilized against Galactic tides by a surrounding dark matter subhalo. 
 \item[iii)] The evolution of luminosity and size of such objects is very sensitive to the (unknown) initial binding energy distribution of a stellar tracer within its surrounding dark matter halo. Stellar systems with near-identical surface brightness profiles may show substantial differences in their binding energy distribution. 
 \item[iv)] In contrast, the tidal evolution of half-light radius and velocity dispersion only weakly depends on the initial stellar energy distribution.
 \item[v)] Velocity dispersion measurements provide a robust criterion to distinguish self-gravitating stellar systems from dark matter-dominated micro galaxies. For micro galaxies with half-light radii of $1 \lesssim \rh/\pc \lesssim 100$, we predict (virial) velocity dispersions between several hundred $\mathrm{m}\,\mathrm{s}^{-1}$ and a few $\kms$. Resolving such dispersions for faint stellar systems represents a considerable challenge. Multi-epoch spectroscopy will be required to accurately account for the contribution of binary stars to the observed dispersions. 
 \item[vi)] Dark matter-dominated micro galaxies have larger mean densities than their self-gravitating counterparts of equal luminosity. For systems where the available velocity measurements are sufficiently accurate to constrain their orbits in the Milky way, but not sufficiently precise to resolve their internal dynamics, the systems' susceptibility to tidal mass loss can serve as an alternative criterion to distinguish globular clusters from micro galaxies.
\end{itemize}

The discovery of dark matter-dominated micro galaxies would inform us about the ability of dark matter subhaloes to survive in the Milky Way tidal field, and at the same time, suggest that stars can populate the most-bound energy states of a halo. Hence, micro galaxies would offer equally insight in the physical properties of dark matter subhaloes, and the baryonic processes that shape the galaxies embedded therein.

\bigskip

\subsection*{Acknowledgements}
RE and RI acknowledge funding from the European Research Council (ERC) under the European Unions Horizon 2020 research and innovation programme (grant agreement number 834148). RE and MW acknowledge support from the National Science Foundation (NSF) grants AST-2206046 and AST-1909584.

\bigskip

\appendix
\section{Comparison of the Scatter in the Initial Conditions Against Cosmological Simulations}
\label{appendix:scatter}
In this section, we show that our initial conditions are consistent with the distribution of circular velocity profiles measured for dark matter haloes in cosmological $N$-body simulations.  

As initial conditions to our modelling framework, we assume NFW density profiles (Eq.~\ref{eq:NFW}) with halo concentrations that follow the \citet{Ludlow2016} concentration-mass-redshift relation (see Sec.~\ref{sec:tidal_evo_model}). We take into account a (logarithmic) scatter of $0.15\,\dex$ in concentration $c=r_{200}/r_\mathrm{s}$ at a given virial mass $M_{200}$ (for definition see footnote~\ref{footnote:virial}).
The value of $0.15\,\dex$ used in this study is a conservative value motivated by the scatter measured in cosmological dark matter only simulations. \citet{DuttonMaccio2014} find that at $z=0$, the scatter around the concentration--mass relation is well approximated by a log-normal distribution of width $0.11\,\dex$. Similarly, \citet[][see their Fig. 4 and 5]{Moline2017} measure a scatter between $0.10 \sim 0.15\,\dex$ for the ELVIS \citep{Garrison-Kimmel2014} and VL-II \citep{Diemand2008Nature} simulations. At higher redshifts, the expected scatter is slightly lower \citep[see e.g.][Fig. B3]{Ludlow2016}. 

At a fixed value of $M_{200}$, assuming NFW profiles, the scatter in the halo characteristic size $\rmx \approx 2.16 \times r_{200}/c$ is identical to the scatter in concentration. Note that $\rmx$ defines the characteristic size of a halo (and, through $\Vmx$, it's mean density) only in combination with a functional form for the density profile. Cosmological simulations show that the circular velocity profiles of CDM haloes have very flat maxima (see e.g. Fig.~1 in \citealt{NavarroLudlow2010}). Small deviations from the NFW functional form of an individual halo, as well as $N$-body discreteness noise, may result in substantial scatter in the exact location of the maximum of the circular velocity profile even in haloes with otherwise very similar density profiles \citep[see e.g.][]{Knebe2013}. 

\begin{figure}
 \centering
 \includegraphics[width=\columnwidth]{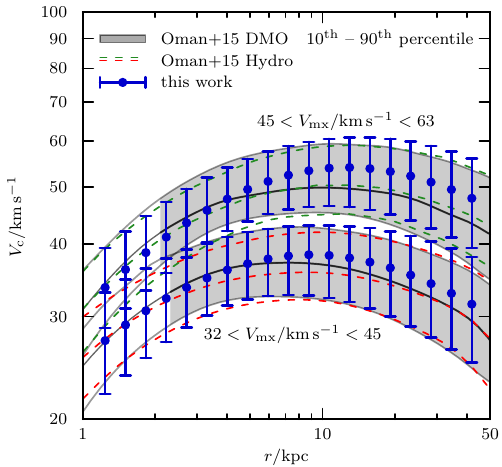}
 \caption{Average circular velocity profiles $\Vc = \left[GM(<r)/r\right]^{1/2}$ for haloes with peak circular velocities $32 < \Vmx/\kms < 45$ and $45 < \Vmx/\kms < 63$. Profiles computed analytically using the same assumptions as adopted in this work are shown as filled blue circles with error bars, spanning from the $10^\mathrm{th}$ to the $90^\mathrm{th}$ percentile of the underlying distribution. Grey shaded bands are taken from \citet[][their Fig.~2]{Oman2015} and show the same range of circular velocity profiles measured in the EAGLE-HR \citep{Schaye2015} and APOSTLE-L2 \citep{Fattahi2016} simulations. The shading is omitted at radii below the \citet{Power2003} convergence radius. }
 \label{fig:AppendixScatterComparison}
\end{figure}

In Fig.\ref{fig:AppendixScatterComparison}, we compare the circular velocity profiles of haloes studied in \citet[][their Fig.~2]{Oman2015} against the equivalent profiles modelled using our simple assumptions. The profiles shown in \citet{Oman2015} are obtained from haloes in the EAGLE-HR \citep{Schaye2015} and APOSTLE-L2 \citep{Fattahi2016} simulations. Note that for halo masses relevant to our study, the profiles discussed in \citet[][]{Oman2015} are virtually identical between the dark matter-only and hydrodynamical runs. We model the concentration-driven scatter in circular velocity profiles by (1) drawing halo virial masses $M_{200}$ from a mass function $\diff N / \diff M_{200} \sim M_{200}^{-1.9}$ \citep[see e.g.][]{Springel2008, Benson2020} (2) drawing halo concentrations $c$ from the \citet{Ludlow2016} mass--concentration relation at redshift $z=0$ assuming a log-normal scatter of $0.15$ dex, (3) computing NFW circular velocity profiles for the sampled values of $M_{200}$ and $c$ and (4) binning the circular velocity profiles using the same bins in $\Vmx$ as adopted in \citet{Oman2015}. 
We find excellent agreement between the $N$-body circular velocity profiles of \citet{Oman2015} and our model, shown respectively as grey bands and filled blue circles with error bars in Fig.~\ref{fig:AppendixScatterComparison}.

\section{Formation Redshift and its Effect on the Structure and the Kinematics of Tidally Limited Dwarfs}
\label{Appendix:FormationRedshift}
The characteristic density of subhaloes is related to their collapse redshift: on average, subhaloes that collapsed earlier also have larger characteristic densities, reflecting the larger mean density of the universe at collapse. 
Similarly, the threshold halo mass for star formation shows a redshift-dependence, and was lower at higher redshifts than it is today. 
In this section, we discuss the observational consequences of the interplay between these two redshift-dependent systematics.

\begin{figure}
 \centering
 \includegraphics[width=\columnwidth]{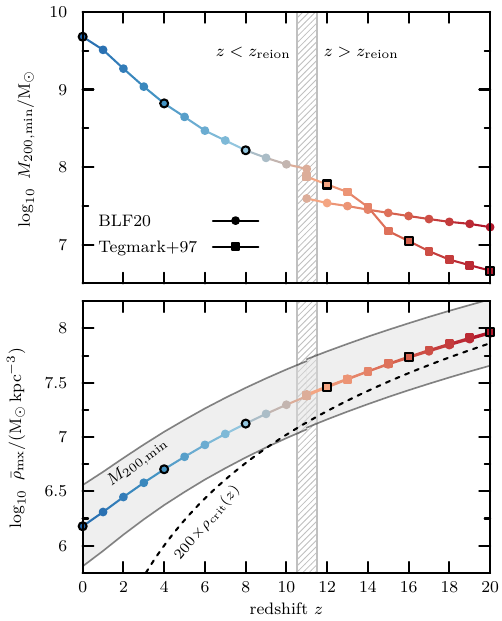}
 \caption{\textit{Top panel:} Virial mass threshold for star formation as a function of redshift, adapted from \citet{Pereira-Wilson2023}. Halo masses computed using the \citet{Benitez-Llambay2020} critical mass model are shown as filled circles. For redshifts $z\ge 11$, we also show the halo mass constraints for hydrogen cooling as in \citet{Tegmark1997} (filled squares). For both models, the threshold halo mass decreases with increasing redshift.  \textit{Bottom panel:} characteristic density $\rho_\mathrm{mx} = \Mmx / (4\pi/3~\rmx^3)$ for haloes with virial mass as in the top panel, using the \citet{Ludlow2016} concentration--mass--redshift relation ($\pm0.15\,\dex$ scatter in concentration, shaded band) . The characteristic density of haloes at the threshold mass for star formation monotonously increases with redshift. For reference, a dashed curve shows a density of $200\times$ the critical density of the universe (see footnote~\ref{footnote:virial}).   }
 \label{fig:m200min}
\end{figure}

In the top panel of Fig.~\ref{fig:m200min}, we show the typical threshold (virial) mass for star formation computed using the \citet{Benitez-Llambay2020} critical mass model, as well for redshifts prior to reionization using the hydrogen cooling limit as in \citet{Tegmark1997}. For both models, the threshold halo mass decreases as redshift increases. Combining this theshold virial mass with halo concentrations computed from the \citet{Ludlow2016} concentration--mass--redshift relation allows to calculate the evolution of the characteristic density $\rho_\mathrm{mx} = \Mmx / (4\pi/3~\rmx^3)$ of the least-massive star forming haloes. This evolution is shown in the bottom panel of Fig.~\ref{fig:m200min}: even though the threshold mass for star formation decreases with redshift, the characteristic density of star forming haloes monotonically increases with redshift. Note that the (rather flat) concentration--mass relation at high redshifts compresses the difference in characteristic density between the \citet{Benitez-Llambay2020} and \citet{Tegmark1997} models. The shaded band corresponds to a scatter in concentration of $\pm0.15\,\dex$.

\begin{figure}
 \centering
 \includegraphics[width=\columnwidth]{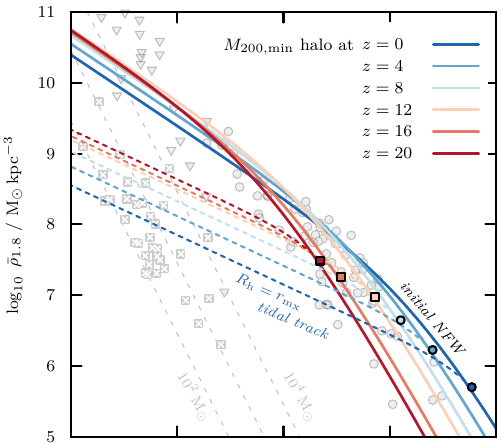}
 \includegraphics[width=\columnwidth]{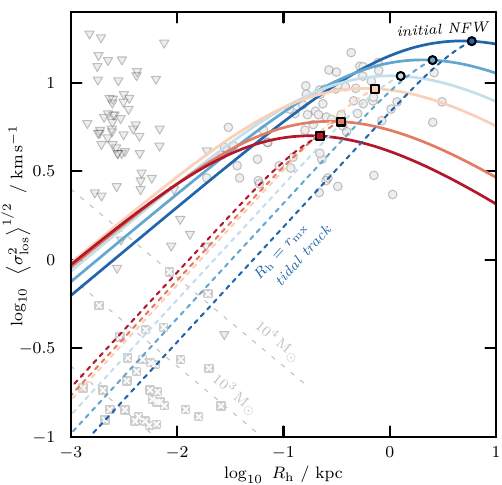}
 \caption{\textit{Top panel:} like Fig.~\ref{fig:data_vs_model_overview}d, showing the average density $\rho_{1.8}$ of a stellar tracer (\texttt{exp3D}, see Sec.~\ref{subsubsec:exp3D}) embedded in a dark matter halo. The halo masses are chosen to match the redshift-dependent threshold mass for star formation as shown in Fig.~\ref{fig:m200min}. Solid curves show $\rho_{1.8}$ for stellar tracers embedded in NFW haloes at redshfits $z=0,\, \dots, 20$. Dashed curves show the tidal evolution of a tidally limited stellar tracer ($\Rh = \rmx$), which defines a lower bound on $\rho_{1.8}$ for a given half-light radius $\Rh$ and redshift $z$. \textit{Bottom panel:} like Fig.~\ref{fig:data_vs_model_overview}b, showing the luminosity-averaged velocity dispersion, but using the same threshold halo masses as in the top panel. For a given half-light radius, a tidally limited stellar system that formed at a higher redshift is expected to have a larger velocity dispersion than a system that formed at a lower redshift.}
 \label{fig:sigma_density_redshift}
\end{figure}

This increase of the characteristic density has direct observable consequences for tidally limited systems. Fig.~\ref{fig:sigma_density_redshift} is identical in structure to  Fig.~\ref{fig:data_vs_model_overview} (b) and (d) and shows the average density $\rho_{1.8}$ (top panel) and the luminosity-averaged velocity dispersion $\langle \sigmalos^2 \rangle^{1/2}$ (bottom panel) of exponential (\texttt{exp3D}) stellar tracers. Here, we choose halo masses corresponding to the minimum mass for star formation at redshifts $z=0,\dots,20$, with matching average concentration for the respective redshift. 

Solid curves show $\rho_{1.8}$ (top panel) and $\langle \sigmalos^2 \rangle^{1/2}$ (bottom panel) for stellar tracers with a half-light radius $\Rh$ embedded in the (initial) NFW halo at different redshifts. Dashed curves of matching colour show the tidal evolution for a tidally limited stellar tracer ($\Rh = \rmx$), delimiting the minimum density (and velocity dispersion) of a tidally stripped system for a given half-light radius $\Rh$ and formation redshift $z$. Tidally limited stellar tracers embedded in progenitor haloes that collapsed at earlier (i.e., higher) redshift are expected to have larger average densities and larger average velocity dispersions at fixed $\Rh$ than their counterparts that formed at lower redshift.

\section{Asymptotic Slopes of dSph Surface Brightness Profiles}
\label{Appendix:ObservarionalConstraints}
In Sec.~\ref{sec:cuspy_vs_cored_tracks}, we discuss the effect of different stellar binding energy distributions on the tidal evolution of dwarf galaxy structural parameters. 
Assuming a stellar tracer deeply embedded in the power-law cusp of an NFW potential, spherical symmetry, and isotropic kinematics, the binding energy distribution can be estimated from the observed surface brightness profile. In this appendix, we broadly discuss how the stellar binding energies shape the observed surface brightness profile (Sec.~\ref{appendix:monoenergy}), and compare the surface brightness profiles underlying the present work against a selection of observed profiles of Milky Way satellites (Sec.~\ref{appendix:observations}). 

\subsection{Illustration of the Connection between Binding Energy Distribution and Surface Brightness Profile}
\label{appendix:monoenergy}
To gain some analytical insight into how the distribution of stellar binding energies shapes the stellar surface brightness profile, consider a mono-energetic stellar distribution function,
\begin{equation}
 f_\star(E) = \delta(E-E_\star)~,
\end{equation}
centred on $E_\star$. The (spherical) stellar density profile is readily obtained by integrating over velocity space, assuming isotropic kinematics,
\begin{equation}
 \rho_\star(r) \propto \!\int\displaylimits_{\text{all }v}\! \diff v ~v^2  \delta \left(  \frac{v^2}{2} + \Phi(r) -E_\star \right) \propto \sqrt{ E_\star - \Phi(r) }
\end{equation}
for $r$ so that $E_\star \geq \Phi(r)$, otherwise $\rho_\star(r) = 0$. This mono-energetic density profile is cored ($\rho_\star(r) \rightarrow\,$const for $r\rightarrow0$) for all potentials $\Phi(r)$ with finite central escape velocity. A centrally-divergent surface brightness profile is therefore a clear indication of the binding energy states being populated all the way down to the most-bound state.

\begin{figure}
 \centering
 \includegraphics[width=\columnwidth]{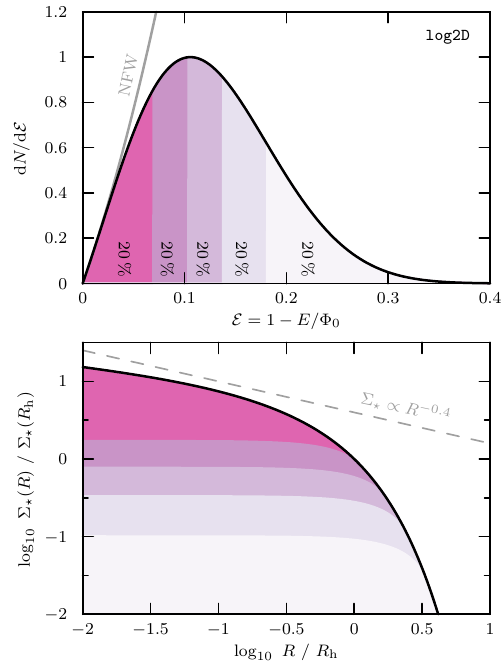}
 \caption{\textit{Top panel:} Binding energy distribution $\diff N / \diff \E$ for the \texttt{log2D} stellar tracer (Sec.~\ref{subsubsec:log2Dstars}) embedded in an NFW halo, for $\Rh/\rmx = 1/16$. The distribution is normalized to $\mathrm{max}(\diff N / \diff \E)=1$. Each shaded energy band contributes $20$ per cent to the total luminosity. \textit{Bottom panel:} Surface brightness profile of the \texttt{log2D} stellar tracer, with shaded regions illustrating the contribution to the total luminosity of the energy bands highlighted in the top panel. All energy bands with a lower bound in the binding energy produce a cored surface brightness profile. The logarithmic cusp in the surface brightness profile is caused by the energy band that includes the most-bound state.}
 \label{fig:energybands}
\end{figure}

We now extend this picture to the contribution of bands in energy to the surface brightness profile. The top panel of Fig.~\ref{fig:energybands} shows the distribution of binding energies of a stellar tracer with a \texttt{log2D} surface brightness profile (see Sec.~\ref{subsubsec:log2Dstars} for definition) embedded in an NFW halo, for $\Rh/\rmx = 1/16$. Bands of binding energy are colour-coded, each band containing 20 per cent of the total luminosity. Note that towards the most-bound energies, the stellar energy distribution is asymptotic to the dark matter (NFW) distribution.
The bottom panel of Fig.~\ref{fig:energybands} shows the corresponding surface brightness profile, colour-coded by the respective contribution of each energy band. All energy bands with a lower bound in the binding energy produce a cored surface brightness profile. The logarithmic cusp of the total surface brightness profile results from the energy band that includes the most-bound state. To guide the eye, a power-law with slope of $\diff \ln \Sigma_\star / \diff \ln R = -0.4$ is shown as a grey dashed curve: In the inner regions of the dwarf, a surface brightness profile where stars follow the dark matter energetically remains very shallow.

\begin{figure}
 \centering
 \includegraphics[width=\columnwidth]{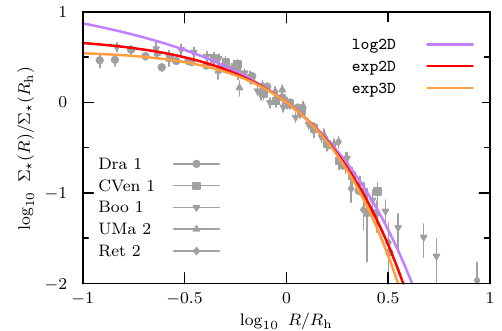}
 \caption{Comparison of the stacked and normalized surface brightness profiles of selected Milky Way satellites against the \texttt{exp2D}~(\gnuplotline{red}), \texttt{exp3D}~(\gnuplotline{tan1}) and \texttt{log2D}~(\gnuplotline{purple}) models (Sec.~\ref{sec:cuspy_vs_cored_tracks}) studied in this work. The surface brightness profiles are computed from star counts available through the DECALS survey \citep{decals19} with photometric selection as in \citet{MoskowitzWalker2020}. The cored \texttt{exp2D} and \texttt{exp3D} appear to better describe the data for Dra~1 and Boo~1 than the logarithmically diverging \texttt{log2D} profile. For CVen~1, UMa~2, Ret~2, all profiles are consistent with the available data.}
 \label{fig:surface_brightness_vs_data}
\end{figure}

\begin{figure}
 \centering
 \includegraphics[width=\columnwidth]{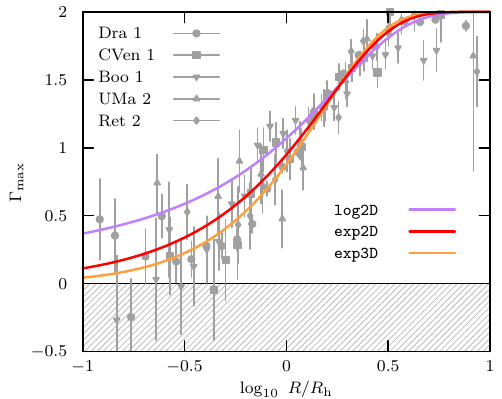}
 \caption{Maximum inner asymptotic slope $\Gamma_\mathrm{max} = - \diff \ln \Sigma_\star(R) / \diff \ln R$ (Eq.~\ref{eq:Gammamax}) of the surface brightness profiles of selected Milky Way satellites, consistent with the enclosed luminosity $L(<R)$ and the local surface brightness $\Sigma_\star(R)$ at a given radius. The available data for the dwarfs shown above constrain the inner slope to $\Gamma_\mathrm{max} \lesssim 0.4$, leaving room for a ``cuspy'' (sub)component to the surface brightness profile. Solid curves show the constraints on $\Gamma_\mathrm{max}$ that could be obtained from surface brightness profiles that (exactly) follow the \texttt{exp2D}, \texttt{exp3D} and \texttt{log2D} models.}
  \label{fig:gammaMax}
\end{figure}

\subsection{Constraints on the Central Asymptotic Slope from Observations}
\label{appendix:observations}
In this section, we compare the surface brightness profiles used in the present study against available observational data of selected Milky Way dwarf galaxies. The observational situation constitutes a conundrum. For the brightest dwarfs with well-measured surface brightness profiles, stellar feedback has likely affected both the dark matter- and stellar distributions, causing central constant-density cores in the dark matter profiles \citep{Pontzen2012, Onorbe2015}. Density cores in turn render the dwarfs vulnerable to tidal mass loss and facilitate their full tidal disruption \citep{Penarrubia2010,ENPFI2023}, preventing them from becoming tidally stripped ``micro galaxies''. Faint dwarfs with few stellar tracers on the other hand may have CDM haloes that are virtually unaffected by their embedded stellar components \citep{Penarrubia2012}, but the low number of available stars in these objects limits to what degree the central slopes of the surface brightness profiles can be constrained observationally. 

As an example, we select five Milky Way satellites spanning a broad range of luminosities: Dra~1, CVen~2 (both with $L \approx 3 \times 10^5\,\Lsol$), Boo~1 ($\approx 2 \times 10^4\,\Lsol$), UMa~2 ($\approx 4 \times 10^3\,\Lsol$) and Ret~2 ($\approx 1 \times 10^3\,\Lsol$). 
We compare the \texttt{exp2D}, \texttt{exp3D} and \texttt{log2D} stellar distributions as defined in Sec.~\ref{sec:cuspy_vs_cored_tracks} against star counts available through the DECALS survey \citep{decals19} for a photometric selection as in \citet{MoskowitzWalker2020}.

Fig.~\ref{fig:surface_brightness_vs_data} shows the surface brightness profiles of the five Milky Way satellites listed above, after fitting and subtracting a uniform background component. For each dwarf, we normalize the projected radii $R$ by the respective half-light radius $\Rh$ of a fitted 2D exponential (\texttt{exp2D}) model. Similarly, the surface brightness of each dwarf is normalized by $\Sigma_\star(\Rh)$ of the fitted 2D exponential. Coloured lines show the surface brightness profiles corresponding to the \texttt{exp2D}, \texttt{exp3D} and \texttt{log2D} models.
For Draco and Bootes 1, the \texttt{exp2D}, \texttt{exp3D} appear to describe the measured profiles better than the \texttt{log2D} model. For the other objects shown, all models provide a reasonable description of the data at those radii where data is available.

A necessary condition for tidally stripped ``micro galaxies'' to exist is that at least a (sub)component of the stellar binding energy distribution extends to the most-bound state. For spherical systems with isotropic kinematics, this translates to a cuspy (sub)component to the surface brightness profile. To address to what extent the observed surface brightness profiles can constrain the presence (or absence) of a central cuspy component, we adopt an approach originally introduced in \citet{Navarro2004} to constrain the inner asymptotic slopes of dark matter (3D) density profiles. Following the same reasoning, but using projected quantities, we aim to compute the maximum asymptotic power-law slope of the surface brightness profile $ \Gamma_\mathrm{max} \equiv - \diff \ln \Sigma_\star / \diff \ln R$ for $R \rightarrow 0$.
We find that the value $\Gamma_\mathrm{max}$ that is compatible with (1) the observed surface brightness $\Sigma_\star(R)$ at radius $R$, (2) the enclosed luminosity $L(<R)$, and (3) a surface brightness profile that is either a power law, or flattens off monotonically towards the centre,  is given by
\begin{equation}
 \Gamma_\mathrm{max} = 2\left[1 - \Sigma_\star(R) \,/\, \bar \Sigma_\star(<R)  \right] ~,
 \label{eq:Gammamax}
\end{equation}
where $\bar \Sigma_\star(<R)  \equiv L(<R) / (\pi R^2)$ is the average surface brightness enclosed within the radius $R$.
In Fig.~\ref{fig:gammaMax}, we show the value of $\Gamma_\mathrm{max}$ as constrained by five example Milky Way dwarf galaxies.
This simple analysis suggests that central cusps in the surface brightness profile with an asymptotic power-law slope of $\Gamma_\mathrm{max} \approx 0.4$ are consistent with the available data, allowing in principle for the existence of a (sub)component of the stellar energy distribution that extends all the way to the most-bound state.

\bigskip

\bibliographystyle{aasjournal}
\bibliography{micro}

\begin{thebibliography}{}
\expandafter\ifx\csname natexlab\endcsname\relax\def\natexlab#1{#1}\fi
\providecommand{\url}[1]{\href{#1}{#1}}
\providecommand{\dodoi}[1]{doi:~\href{http://doi.org/#1}{\nolinkurl{#1}}}
\providecommand{\doeprint}[1]{\href{http://ascl.net/#1}{\nolinkurl{http://ascl.net/#1}}}
\providecommand{\doarXiv}[1]{\href{https://arxiv.org/abs/#1}{\nolinkurl{https://arxiv.org/abs/#1}}}

\bibitem[{{Abramowitz} \& {Stegun}(1972)}]{AbramowitzStegun1972}
{Abramowitz}, M., \& {Stegun}, I.~A. 1972, {Handbook of Mathematical Functions}

\bibitem[{{Ahvazi} {et~al.}(2024){Ahvazi}, {Benson}, {Sales}, {Nadler},
  {Weerasooriya}, {Du}, \& {Bovill}}]{Ahvazi2023}
{Ahvazi}, N., {Benson}, A., {Sales}, L.~V., {et~al.} 2024, \mnras, 529, 3387,
  \dodoi{10.1093/mnras/stae761}

\bibitem[{{Amorisco}(2021)}]{Amorisco2021}
{Amorisco}, N.~C. 2021, arXiv e-prints, arXiv:2111.01148.
\newblock \doarXiv{2111.01148}

\bibitem[{{Amorisco} {et~al.}(2013){Amorisco}, {Agnello}, \&
  {Evans}}]{AmoriscoAgnelloEvans2013}
{Amorisco}, N.~C., {Agnello}, A., \& {Evans}, N.~W. 2013, \mnras, 429, L89,
  \dodoi{10.1093/mnrasl/sls031}

\bibitem[{{Amorisco} \& {Evans}(2012)}]{Amorisco2012}
{Amorisco}, N.~C., \& {Evans}, N.~W. 2012, \mnras, 419, 184,
  \dodoi{10.1111/j.1365-2966.2011.19684.x}

\bibitem[{{Battaglia} \& {Nipoti}(2022)}]{Battaglia2022}
{Battaglia}, G., \& {Nipoti}, C. 2022, Nature Astronomy, 6, 659,
  \dodoi{10.1038/s41550-022-01638-7}

\bibitem[{{Battaglia} {et~al.}(2022){Battaglia}, {Taibi}, {Thomas}, \&
  {Fritz}}]{Battaglia2022_Gaia}
{Battaglia}, G., {Taibi}, S., {Thomas}, G.~F., \& {Fritz}, T.~K. 2022, \aap,
  657, A54, \dodoi{10.1051/0004-6361/202141528}

\bibitem[{{Baumgardt} {et~al.}(2009){Baumgardt}, {C{\^o}t{\'e}}, {Hilker},
  {Rejkuba}, {Mieske}, {Djorgovski}, \& {Stetson}}]{Baumgardt2009}
{Baumgardt}, H., {C{\^o}t{\'e}}, P., {Hilker}, M., {et~al.} 2009, \mnras, 396,
  2051, \dodoi{10.1111/j.1365-2966.2009.14932.x}

\bibitem[{{Benitez-Llambay} \& {Frenk}(2020)}]{Benitez-Llambay2020}
{Benitez-Llambay}, A., \& {Frenk}, C. 2020, \mnras, 498, 4887,
  \dodoi{10.1093/mnras/staa2698}

\bibitem[{{Ben{\'\i}tez-Llambay} {et~al.}(2019){Ben{\'\i}tez-Llambay}, {Frenk},
  {Ludlow}, \& {Navarro}}]{Benitez-Llambay2019}
{Ben{\'\i}tez-Llambay}, A., {Frenk}, C.~S., {Ludlow}, A.~D., \& {Navarro},
  J.~F. 2019, \mnras, 488, 2387, \dodoi{10.1093/mnras/stz1890}

\bibitem[{{Benson}(2020)}]{Benson2020}
{Benson}, A.~J. 2020, \mnras, 493, 1268, \dodoi{10.1093/mnras/staa341}

\bibitem[{{Binney} \& {Tremaine}(1987)}]{BT87}
{Binney}, J., \& {Tremaine}, S. 1987, {Galactic dynamics}

\bibitem[{{Borukhovetskaya} {et~al.}(2022){Borukhovetskaya}, {Navarro},
  {Errani}, \& {Fattahi}}]{Borukhovetskaya2022}
{Borukhovetskaya}, A., {Navarro}, J.~F., {Errani}, R., \& {Fattahi}, A. 2022,
  \mnras, 512, 5247, \dodoi{10.1093/mnras/stac653}

\bibitem[{{Bose} {et~al.}(2020){Bose}, {Deason}, {Belokurov}, \&
  {Frenk}}]{Bose2020}
{Bose}, S., {Deason}, A.~J., {Belokurov}, V., \& {Frenk}, C.~S. 2020, \mnras,
  495, 743, \dodoi{10.1093/mnras/staa1199}

\bibitem[{{Bovill} \& {Ricotti}(2009)}]{BovillRicotti2009}
{Bovill}, M.~S., \& {Ricotti}, M. 2009, \apj, 693, 1859,
  \dodoi{10.1088/0004-637X/693/2/1859}

\bibitem[{{Bruce} {et~al.}(2023){Bruce}, {Li}, {Pace}, {Heiger}, {Song}, \&
  {Simon}}]{Bruce2023}
{Bruce}, J., {Li}, T.~S., {Pace}, A.~B., {et~al.} 2023, \apj, 950, 167,
  \dodoi{10.3847/1538-4357/acc943}

\bibitem[{{Buist} \& {Helmi}(2016)}]{buisthelmi16}
{Buist}, H.~J.~T., \& {Helmi}, A. 2016, \aap, 589, C3,
  \dodoi{10.1051/0004-6361/201323059e}

\bibitem[{{Bullock} {et~al.}(2000){Bullock}, {Kravtsov}, \&
  {Weinberg}}]{Bullock2000}
{Bullock}, J.~S., {Kravtsov}, A.~V., \& {Weinberg}, D.~H. 2000, \apj, 539, 517,
  \dodoi{10.1086/309279}

\bibitem[{{Burkert}(2000)}]{Burkert2000}
{Burkert}, A. 2000, \apjl, 534, L143, \dodoi{10.1086/312674}

\bibitem[{{Cautun} \& {Frenk}(2017)}]{Cautun2017}
{Cautun}, M., \& {Frenk}, C.~S. 2017, \mnras, 468, L41,
  \dodoi{10.1093/mnrasl/slx025}

\bibitem[{{Cerny} {et~al.}(2023{\natexlab{a}}){Cerny},
  {Mart{\'\i}nez-V{\'a}zquez}, {Drlica-Wagner}, {Pace}, {Mutlu-Pakdil}, {Li},
  {Riley}, {Crnojevi{\'c}}, {Bom}, {Carballo-Bello}, {Carlin}, {Chiti}, {Choi},
  {Collins}, {Darragh-Ford}, {Ferguson}, {Geha}, {Mart{\'\i}nez-Delgado},
  {Massana}, {Mau}, {Medina}, {Mu{\~n}oz}, {Nadler}, {No{\"e}l}, {Olsen},
  {Pieres}, {Sakowska}, {Simon}, {Stringfellow}, {Tollerud}, {Vivas}, {Walker},
  {Wechsler}, \& {Delve Collaboration}}]{Cerny22}
{Cerny}, W., {Mart{\'\i}nez-V{\'a}zquez}, C.~E., {Drlica-Wagner}, A., {et~al.}
  2023{\natexlab{a}}, \apj, 953, 1, \dodoi{10.3847/1538-4357/acdd78}

\bibitem[{{Cerny} {et~al.}(2023{\natexlab{b}}){Cerny}, {Simon}, {Li},
  {Drlica-Wagner}, {Pace}, {Mart{\'\i}nez-V{\'a}zquez}, {Riley},
  {Mutlu-Pakdil}, {Mau}, {Ferguson}, {Erkal}, {Munoz}, {Bom}, {Carlin},
  {Carollo}, {Choi}, {Ji}, {Manwadkar}, {Mart{\'\i}nez-Delgado}, {Miller},
  {No{\"e}l}, {Sakowska}, {Sand}, {Stringfellow}, {Tollerud}, {Vivas},
  {Carballo-Bello}, {Hernandez-Lang}, {James}, {Nidever}, {Castellon}, {Olsen},
  {Zenteno}, \& {Delve Collaboration}}]{Cerny23}
{Cerny}, W., {Simon}, J.~D., {Li}, T.~S., {et~al.} 2023{\natexlab{b}}, \apj,
  942, 111, \dodoi{10.3847/1538-4357/aca1c3}

\bibitem[{{Chiti} {et~al.}(2021){Chiti}, {Frebel}, {Simon}, {Erkal}, {Chang},
  {Necib}, {Ji}, {Jerjen}, {Kim}, \& {Norris}}]{Chiti2021}
{Chiti}, A., {Frebel}, A., {Simon}, J.~D., {et~al.} 2021, Nature Astronomy, 5,
  392, \dodoi{10.1038/s41550-020-01285-w}

\bibitem[{{Col{\'\i}n} {et~al.}(2002){Col{\'\i}n}, {Avila-Reese}, {Valenzuela},
  \& {Firmani}}]{Colin2002}
{Col{\'\i}n}, P., {Avila-Reese}, V., {Valenzuela}, O., \& {Firmani}, C. 2002,
  \apj, 581, 777, \dodoi{10.1086/344259}

\bibitem[{{Collins} {et~al.}(2020){Collins}, {Tollerud}, {Rich}, {Ibata},
  {Martin}, {Chapman}, {Gilbert}, \& {Preston}}]{Collins2020}
{Collins}, M. L.~M., {Tollerud}, E.~J., {Rich}, R.~M., {et~al.} 2020, \mnras,
  491, 3496, \dodoi{10.1093/mnras/stz3252}

\bibitem[{{Collins} {et~al.}(2021){Collins}, {Read}, {Ibata}, {Rich}, {Martin},
  {Pe{\~n}arrubia}, {Chapman}, {Tollerud}, \& {Weisz}}]{Collins2021}
{Collins}, M. L.~M., {Read}, J.~I., {Ibata}, R.~A., {et~al.} 2021, \mnras, 505,
  5686, \dodoi{10.1093/mnras/stab1624}

\bibitem[{{Delos} \& {White}(2023{\natexlab{a}})}]{Delos2023gamma}
{Delos}, M.~S., \& {White}, S. D.~M. 2023{\natexlab{a}}, \jcap, 2023, 008,
  \dodoi{10.1088/1475-7516/2023/10/008}

\bibitem[{{Delos} \& {White}(2023{\natexlab{b}})}]{DelosWhite2023}
---. 2023{\natexlab{b}}, \mnras, 518, 3509, \dodoi{10.1093/mnras/stac3373}

\bibitem[{{Dey} {et~al.}(2019){Dey}, {Schlegel}, {Lang}, {Blum}, {Burleigh},
  {Fan}, {Findlay}, {Finkbeiner}, {Herrera}, {Juneau}, {Landriau}, {Levi},
  {McGreer}, {Meisner}, {Myers}, {Moustakas}, {Nugent}, {Patej}, {Schlafly},
  {Walker}, {Valdes}, {Weaver}, {Y{\`e}che}, {Zou}, {Zhou}, {Abareshi},
  {Abbott}, {Abolfathi}, {Aguilera}, {Alam}, {Allen}, {Alvarez}, {Annis},
  {Ansarinejad}, {Aubert}, {Beechert}, {Bell}, {BenZvi}, {Beutler}, {Bielby},
  {Bolton}, {Brice{\~n}o}, {Buckley-Geer}, {Butler}, {Calamida}, {Carlberg},
  {Carter}, {Casas}, {Castander}, {Choi}, {Comparat}, {Cukanovaite}, {Delubac},
  {DeVries}, {Dey}, {Dhungana}, {Dickinson}, {Ding}, {Donaldson}, {Duan},
  {Duckworth}, {Eftekharzadeh}, {Eisenstein}, {Etourneau}, {Fagrelius},
  {Farihi}, {Fitzpatrick}, {Font-Ribera}, {Fulmer}, {G{\"a}nsicke},
  {Gaztanaga}, {George}, {Gerdes}, {Gontcho}, {Gorgoni}, {Green}, {Guy},
  {Harmer}, {Hernandez}, {Honscheid}, {Huang}, {James}, {Jannuzi}, {Jiang},
  {Joyce}, {Karcher}, {Karkar}, {Kehoe}, {Kneib}, {Kueter-Young}, {Lan},
  {Lauer}, {Le Guillou}, {Le Van Suu}, {Lee}, {Lesser}, {Perreault Levasseur},
  {Li}, {Mann}, {Marshall}, {Mart{\'\i}nez-V{\'a}zquez}, {Martini}, {du Mas des
  Bourboux}, {McManus}, {Meier}, {M{\'e}nard}, {Metcalfe},
  {Mu{\~n}oz-Guti{\'e}rrez}, {Najita}, {Napier}, {Narayan}, {Newman}, {Nie},
  {Nord}, {Norman}, {Olsen}, {Paat}, {Palanque-Delabrouille}, {Peng},
  {Poppett}, {Poremba}, {Prakash}, {Rabinowitz}, {Raichoor}, {Rezaie},
  {Robertson}, {Roe}, {Ross}, {Ross}, {Rudnick}, {Safonova}, {Saha},
  {S{\'a}nchez}, {Savary}, {Schweiker}, {Scott}, {Seo}, {Shan}, {Silva},
  {Slepian}, {Soto}, {Sprayberry}, {Staten}, {Stillman}, {Stupak}, {Summers},
  {Sien Tie}, {Tirado}, {Vargas-Maga{\~n}a}, {Vivas}, {Wechsler}, {Williams},
  {Yang}, {Yang}, {Yapici}, {Zaritsky}, {Zenteno}, {Zhang}, {Zhang}, {Zhou}, \&
  {Zhou}}]{decals19}
{Dey}, A., {Schlegel}, D.~J., {Lang}, D., {et~al.} 2019, \aj, 157, 168,
  \dodoi{10.3847/1538-3881/ab089d}

\bibitem[{{Diakogiannis} {et~al.}(2017){Diakogiannis}, {Lewis}, {Ibata},
  {Guglielmo}, {Kafle}, {Wilkinson}, \& {Power}}]{Diakogiannis2017}
{Diakogiannis}, F.~I., {Lewis}, G.~F., {Ibata}, R.~A., {et~al.} 2017, \mnras,
  470, 2034, \dodoi{10.1093/mnras/stx1219}

\bibitem[{{Diemand} {et~al.}(2008){Diemand}, {Kuhlen}, {Madau}, {Zemp},
  {Moore}, {Potter}, \& {Stadel}}]{Diemand2008Nature}
{Diemand}, J., {Kuhlen}, M., {Madau}, P., {et~al.} 2008, \nat, 454, 735,
  \dodoi{10.1038/nature07153}

\bibitem[{{D'Onghia} {et~al.}(2010){D'Onghia}, {Springel}, {Hernquist}, \&
  {Keres}}]{DOnghia10}
{D'Onghia}, E., {Springel}, V., {Hernquist}, L., \& {Keres}, D. 2010, \apj,
  709, 1138, \dodoi{10.1088/0004-637X/709/2/1138}

\bibitem[{{Dutton} \& {Macci{\`o}}(2014)}]{DuttonMaccio2014}
{Dutton}, A.~A., \& {Macci{\`o}}, A.~V. 2014, \mnras, 441, 3359,
  \dodoi{10.1093/mnras/stu742}

\bibitem[{{Errani} \& {Navarro}(2021)}]{EN21}
{Errani}, R., \& {Navarro}, J.~F. 2021, \mnras, 505, 18,
  \dodoi{10.1093/mnras/stab1215}

\bibitem[{{Errani} {et~al.}(2022){Errani}, {Navarro}, {Ibata}, \&
  {Pe{\~n}arrubia}}]{ENIP2022}
{Errani}, R., {Navarro}, J.~F., {Ibata}, R., \& {Pe{\~n}arrubia}, J. 2022,
  \mnras, 511, 6001, \dodoi{10.1093/mnras/stac476}

\bibitem[{{Errani} {et~al.}(2023){Errani}, {Navarro}, {Pe{\~n}arrubia},
  {Famaey}, \& {Ibata}}]{ENPFI2023}
{Errani}, R., {Navarro}, J.~F., {Pe{\~n}arrubia}, J., {Famaey}, B., \& {Ibata},
  R. 2023, \mnras, 519, 384, \dodoi{10.1093/mnras/stac3499}

\bibitem[{{Errani} {et~al.}(2024){Errani}, {Navarro}, {Smith}, \&
  {McConnachie}}]{ENSM24}
{Errani}, R., {Navarro}, J.~F., {Smith}, S. E.~T., \& {McConnachie}, A.~W.
  2024, \apj, 965, 20, \dodoi{10.3847/1538-4357/ad2267}

\bibitem[{{Errani} \& {Pe{\~n}arrubia}(2020)}]{EP20}
{Errani}, R., \& {Pe{\~n}arrubia}, J. 2020, \mnras, 491, 4591,
  \dodoi{10.1093/mnras/stz3349}

\bibitem[{{Errani} {et~al.}(2017){Errani}, {Pe{\~n}arrubia}, {Laporte}, \&
  {G{\'o}mez}}]{EPLG17}
{Errani}, R., {Pe{\~n}arrubia}, J., {Laporte}, C.~F.~P., \& {G{\'o}mez}, F.~A.
  2017, \mnras, 465, L59, \dodoi{10.1093/mnrasl/slw211}

\bibitem[{{Errani} {et~al.}(2018){Errani}, {Pe{\~n}arrubia}, \&
  {Walker}}]{EPW18}
{Errani}, R., {Pe{\~n}arrubia}, J., \& {Walker}, M.~G. 2018, \mnras, 481, 5073,
  \dodoi{10.1093/mnras/sty2505}

\bibitem[{{Fattahi} {et~al.}(2018){Fattahi}, {Navarro}, {Frenk}, {Oman},
  {Sawala}, \& {Schaller}}]{Fattahi2018}
{Fattahi}, A., {Navarro}, J.~F., {Frenk}, C.~S., {et~al.} 2018, \mnras, 476,
  3816, \dodoi{10.1093/mnras/sty408}

\bibitem[{{Fattahi} {et~al.}(2016){Fattahi}, {Navarro}, {Sawala}, {Frenk},
  {Oman}, {Crain}, {Furlong}, {Schaller}, {Schaye}, {Theuns}, \&
  {Jenkins}}]{Fattahi2016}
{Fattahi}, A., {Navarro}, J.~F., {Sawala}, T., {et~al.} 2016, \mnras, 457, 844,
  \dodoi{10.1093/mnras/stv2970}

\bibitem[{{Fellhauer} {et~al.}(2000){Fellhauer}, {Kroupa}, {Baumgardt}, {Bien},
  {Boily}, {Spurzem}, \& {Wassmer}}]{Fellhauer2000}
{Fellhauer}, M., {Kroupa}, P., {Baumgardt}, H., {et~al.} 2000, NA, 5, 305,
  \dodoi{10.1016/S1384-1076(00)00032-4}

\bibitem[{{Frenk} \& {White}(2012)}]{FrenkWhite2012}
{Frenk}, C.~S., \& {White}, S.~D.~M. 2012, Annalen der Physik, 524, 507,
  \dodoi{10.1002/andp.201200212}

\bibitem[{{Garrison-Kimmel} {et~al.}(2014){Garrison-Kimmel}, {Boylan-Kolchin},
  {Bullock}, \& {Lee}}]{Garrison-Kimmel2014}
{Garrison-Kimmel}, S., {Boylan-Kolchin}, M., {Bullock}, J.~S., \& {Lee}, K.
  2014, \mnras, 438, 2578, \dodoi{10.1093/mnras/stt2377}

\bibitem[{{Gieles} {et~al.}(2021){Gieles}, {Erkal}, {Antonini}, {Balbinot}, \&
  {Pe{\~n}arrubia}}]{Gieles2021}
{Gieles}, M., {Erkal}, D., {Antonini}, F., {Balbinot}, E., \& {Pe{\~n}arrubia},
  J. 2021, Nature Astronomy, 5, 957, \dodoi{10.1038/s41550-021-01392-2}

\bibitem[{{Gnedin}(2000)}]{Gnedin2000}
{Gnedin}, N.~Y. 2000, \apj, 542, 535, \dodoi{10.1086/317042}

\bibitem[{{Green} {et~al.}(2021){Green}, {van den Bosch}, \&
  {Jiang}}]{Green2021_mass_function}
{Green}, S.~B., {van den Bosch}, F.~C., \& {Jiang}, F. 2021, \mnras, 503, 4075,
  \dodoi{10.1093/mnras/stab696}

\bibitem[{{Guo} {et~al.}(2010){Guo}, {White}, {Li}, \&
  {Boylan-Kolchin}}]{Guo2010}
{Guo}, Q., {White}, S., {Li}, C., \& {Boylan-Kolchin}, M. 2010, \mnras, 404,
  1111, \dodoi{10.1111/j.1365-2966.2010.16341.x}

\bibitem[{{Guo} {et~al.}(2011){Guo}, {White}, {Boylan-Kolchin}, {De Lucia},
  {Kauffmann}, {Lemson}, {Li}, {Springel}, \& {Weinmann}}]{Guo2011}
{Guo}, Q., {White}, S., {Boylan-Kolchin}, M., {et~al.} 2011, \mnras, 413, 101,
  \dodoi{10.1111/j.1365-2966.2010.18114.x}

\bibitem[{{Han} {et~al.}(2016){Han}, {Cole}, {Frenk}, \& {Jing}}]{Han2016}
{Han}, J., {Cole}, S., {Frenk}, C.~S., \& {Jing}, Y. 2016, \mnras, 457, 1208,
  \dodoi{10.1093/mnras/stv2900}

\bibitem[{{Harris}(1996)}]{Harris1996}
{Harris}, W.~E. 1996, \aj, 112, 1487, \dodoi{10.1086/118116}

\bibitem[{{Hayashi} {et~al.}(2020){Hayashi}, {Chiba}, \&
  {Ishiyama}}]{Hayashi2020}
{Hayashi}, K., {Chiba}, M., \& {Ishiyama}, T. 2020, \apj, 904, 45,
  \dodoi{10.3847/1538-4357/abbe0a}

\bibitem[{{Hilker}(2006)}]{Hiker2006}
{Hilker}, M. 2006, \aap, 448, 171, \dodoi{10.1051/0004-6361:20054327}

\bibitem[{{Illingworth}(1976)}]{Illingworth1976}
{Illingworth}, G. 1976, \apj, 204, 73, \dodoi{10.1086/154152}

\bibitem[{{Irwin} \& {Hatzidimitriou}(1995)}]{Irwin1995}
{Irwin}, M., \& {Hatzidimitriou}, D. 1995, \mnras, 277, 1354,
  \dodoi{10.1093/mnras/277.4.1354}

\bibitem[{{Ishiyama} {et~al.}(2010){Ishiyama}, {Makino}, \&
  {Ebisuzaki}}]{Ishiyama2010}
{Ishiyama}, T., {Makino}, J., \& {Ebisuzaki}, T. 2010, \apjl, 723, L195,
  \dodoi{10.1088/2041-8205/723/2/L195}

\bibitem[{{Jardel} {et~al.}(2013){Jardel}, {Gebhardt}, {Fabricius}, {Drory}, \&
  {Williams}}]{Jardel2013}
{Jardel}, J.~R., {Gebhardt}, K., {Fabricius}, M.~H., {Drory}, N., \&
  {Williams}, M.~J. 2013, \apj, 763, 91, \dodoi{10.1088/0004-637X/763/2/91}

\bibitem[{{Jensen} {et~al.}(2024){Jensen}, {Hayes}, {Sestito}, {McConnachie},
  {Waller}, {Smith}, {Navarro}, \& {Venn}}]{Jensen2024}
{Jensen}, J., {Hayes}, C.~R., {Sestito}, F., {et~al.} 2024, \mnras, 527, 4209,
  \dodoi{10.1093/mnras/stad3322}

\bibitem[{{Ji} {et~al.}(2021){Ji}, {Koposov}, {Li}, {Erkal}, {Pace}, {Simon},
  {Belokurov}, {Cullinane}, {Da Costa}, {Kuehn}, {Lewis}, {Mackey}, {Shipp},
  {Simpson}, {Zucker}, {Hansen}, {Bland-Hawthorn}, \& {S5
  Collaboration}}]{Ji2021}
{Ji}, A.~P., {Koposov}, S.~E., {Li}, T.~S., {et~al.} 2021, \apj, 921, 32,
  \dodoi{10.3847/1538-4357/ac1869}

\bibitem[{{Jordi} {et~al.}(2009){Jordi}, {Grebel}, {Hilker}, {Baumgardt},
  {Frank}, {Kroupa}, {Haghi}, {C{\^o}t{\'e}}, \& {Djorgovski}}]{Jordi2009}
{Jordi}, K., {Grebel}, E.~K., {Hilker}, M., {et~al.} 2009, \aj, 137, 4586,
  \dodoi{10.1088/0004-6256/137/6/4586}

\bibitem[{{Kelley} {et~al.}(2019){Kelley}, {Bullock}, {Garrison-Kimmel},
  {Boylan-Kolchin}, {Pawlowski}, \& {Graus}}]{Kelley2019}
{Kelley}, T., {Bullock}, J.~S., {Garrison-Kimmel}, S., {et~al.} 2019, \mnras,
  487, 4409, \dodoi{10.1093/mnras/stz1553}

\bibitem[{{Knebe} {et~al.}(2013){Knebe}, {Pearce}, {Lux}, {Ascasibar},
  {Behroozi}, {Casado}, {Moran}, {Diemand}, {Dolag}, {Dominguez-Tenreiro},
  {Elahi}, {Falck}, {Gottl{\"o}ber}, {Han}, {Klypin}, {Luki{\'c}},
  {Maciejewski}, {McBride}, {Merch{\'a}n}, {Muldrew}, {Neyrinck}, {Onions},
  {Planelles}, {Potter}, {Quilis}, {Rasera}, {Ricker}, {Roy}, {Ruiz},
  {Sgr{\'o}}, {Springel}, {Stadel}, {Sutter}, {Tweed}, \& {Zemp}}]{Knebe2013}
{Knebe}, A., {Pearce}, F.~R., {Lux}, H., {et~al.} 2013, \mnras, 435, 1618,
  \dodoi{10.1093/mnras/stt1403}

\bibitem[{{Koposov} {et~al.}(2011){Koposov}, {Gilmore}, {Walker}, {Belokurov},
  {Evans}, {Fellhauer}, {Gieren}, {Geisler}, {Monaco}, {Norris}, {Okamoto},
  {Pe{\~n}arrubia}, {Wilkinson}, {Wyse}, \& {Zucker}}]{Koposov2011}
{Koposov}, S.~E., {Gilmore}, G., {Walker}, M.~G., {et~al.} 2011, \apj, 736,
  146, \dodoi{10.1088/0004-637X/736/2/146}

\bibitem[{{Kravtsov}(2010)}]{Kravtsov2010}
{Kravtsov}, A. 2010, Advances in Astronomy, 2010, 281913,
  \dodoi{10.1155/2010/281913}

\bibitem[{{Kuzma} {et~al.}(2015){Kuzma}, {Da Costa}, {Keller}, \&
  {Maunder}}]{Kuzma2015}
{Kuzma}, P.~B., {Da Costa}, G.~S., {Keller}, S.~C., \& {Maunder}, E. 2015,
  \mnras, 446, 3297, \dodoi{10.1093/mnras/stu2343}

\bibitem[{{Li} {et~al.}(2021){Li}, {Hammer}, {Babusiaux}, {Pawlowski}, {Yang},
  {Arenou}, {Du}, \& {Wang}}]{Li2021_Gaia}
{Li}, H., {Hammer}, F., {Babusiaux}, C., {et~al.} 2021, \apj, 916, 8,
  \dodoi{10.3847/1538-4357/ac0436}

\bibitem[{{Libeskind} {et~al.}(2010){Libeskind}, {Yepes}, {Knebe},
  {Gottl{\"o}ber}, {Hoffman}, \& {Knollmann}}]{Libeskind2010}
{Libeskind}, N.~I., {Yepes}, G., {Knebe}, A., {et~al.} 2010, \mnras, 401, 1889,
  \dodoi{10.1111/j.1365-2966.2009.15766.x}

\bibitem[{{Longeard} {et~al.}(2018){Longeard}, {Martin}, {Starkenburg},
  {Ibata}, {Collins}, {Geha}, {Laevens}, {Rich}, {Aguado}, {Arentsen},
  {Carlberg}, {C{\^o}t{\'e}}, {Hill}, {Jablonka}, {Gonz{\'a}lez Hern{\'a}ndez},
  {Navarro}, {S{\'a}nchez-Janssen}, {Tolstoy}, {Venn}, \&
  {Youakim}}]{Longeard2018}
{Longeard}, N., {Martin}, N., {Starkenburg}, E., {et~al.} 2018, \mnras, 480,
  2609, \dodoi{10.1093/mnras/sty1986}

\bibitem[{{Ludlow} {et~al.}(2016){Ludlow}, {Bose}, {Angulo}, {Wang},
  {Hellwing}, {Navarro}, {Cole}, \& {Frenk}}]{Ludlow2016}
{Ludlow}, A.~D., {Bose}, S., {Angulo}, R.~E., {et~al.} 2016, \mnras, 460, 1214,
  \dodoi{10.1093/mnras/stw1046}

\bibitem[{{Ludlow} {et~al.}(2013){Ludlow}, {Navarro}, {Boylan-Kolchin}, {Bett},
  {Angulo}, {Li}, {White}, {Frenk}, \& {Springel}}]{LudlowNavarro2013}
{Ludlow}, A.~D., {Navarro}, J.~F., {Boylan-Kolchin}, M., {et~al.} 2013, \mnras,
  432, 1103, \dodoi{10.1093/mnras/stt526}

\bibitem[{{Manwadkar} \& {Kravtsov}(2022)}]{Manwadkar2022}
{Manwadkar}, V., \& {Kravtsov}, A.~V. 2022, \mnras, 516, 3944,
  \dodoi{10.1093/mnras/stac2452}

\bibitem[{{Martin} {et~al.}(2016){Martin}, {Geha}, {Ibata}, {Collins},
  {Laevens}, {Bell}, {Rix}, {Ferguson}, {Chambers}, {Wainscoat}, \&
  {Waters}}]{Martin2016_Dra2}
{Martin}, N.~F., {Geha}, M., {Ibata}, R.~A., {et~al.} 2016, \mnras, 458, L59,
  \dodoi{10.1093/mnrasl/slw013}

\bibitem[{{Massari} {et~al.}(2020){Massari}, {Helmi}, {Mucciarelli}, {Sales},
  {Spina}, \& {Tolstoy}}]{Massari2020}
{Massari}, D., {Helmi}, A., {Mucciarelli}, A., {et~al.} 2020, \aap, 633, A36,
  \dodoi{10.1051/0004-6361/201935613}

\bibitem[{{Mateo} {et~al.}(1993){Mateo}, {Olszewski}, {Pryor}, {Welch}, \&
  {Fischer}}]{Mateo1993}
{Mateo}, M., {Olszewski}, E.~W., {Pryor}, C., {Welch}, D.~L., \& {Fischer}, P.
  1993, \aj, 105, 510, \dodoi{10.1086/116449}

\bibitem[{{Mateo}(1998)}]{Mateo1998}
{Mateo}, M.~L. 1998, \araa, 36, 435, \dodoi{10.1146/annurev.astro.36.1.435}

\bibitem[{{Mau} {et~al.}(2020){Mau}, {Cerny}, {Pace}, {Choi}, {Drlica-Wagner},
  {Santana-Silva}, {Riley}, {Erkal}, {Stringfellow}, {Adam{\'o}w}, {Carlin},
  {Gruendl}, {Hernandez-Lang}, {Kuropatkin}, {Li}, {Mart{\'\i}nez-V{\'a}zquez},
  {Morganson}, {Mutlu-Pakdil}, {Neilsen}, {Nidever}, {Olsen}, {Sand},
  {Tollerud}, {Tucker}, {Yanny}, {Zenteno}, {Allam}, {Barkhouse}, {Bechtol},
  {Bell}, {Balaji}, {Crnojevi{\'c}}, {Esteves}, {Ferguson}, {Gallart},
  {Hughes}, {James}, {Jethwa}, {Johnson}, {Kuehn}, {Majewski}, {Mao},
  {Massana}, {McNanna}, {Monachesi}, {Nadler}, {No{\"e}l}, {Palmese},
  {Paz-Chinchon}, {Pieres}, {Sanchez}, {Shipp}, {Simon}, {Soares-Santos},
  {Tavangar}, {van der Marel}, {Vivas}, {Walker}, \& {Wechsler}}]{Mau2020}
{Mau}, S., {Cerny}, W., {Pace}, A.~B., {et~al.} 2020, \apj, 890, 136,
  \dodoi{10.3847/1538-4357/ab6c67}

\bibitem[{{McConnachie}(2012)}]{McConnachie2012}
{McConnachie}, A.~W. 2012, \aj, 144, 4, \dodoi{10.1088/0004-6256/144/1/4}

\bibitem[{{McConnachie} \& {C{\^o}t{\'e}}(2010)}]{McConnachieCote2010}
{McConnachie}, A.~W., \& {C{\^o}t{\'e}}, P. 2010, \apjl, 722, L209,
  \dodoi{10.1088/2041-8205/722/2/L209}

\bibitem[{{Merritt}(1987)}]{Merritt1987}
{Merritt}, D. 1987, \apj, 313, 121, \dodoi{10.1086/164953}

\bibitem[{{Minor} {et~al.}(2019){Minor}, {Pace}, {Marshall}, \&
  {Strigari}}]{Minor2019}
{Minor}, Q.~E., {Pace}, A.~B., {Marshall}, J.~L., \& {Strigari}, L.~E. 2019,
  \mnras, 487, 2961, \dodoi{10.1093/mnras/stz1468}

\bibitem[{{Molin{\'e}} {et~al.}(2017){Molin{\'e}}, {S{\'a}nchez-Conde},
  {Palomares-Ruiz}, \& {Prada}}]{Moline2017}
{Molin{\'e}}, {\'A}., {S{\'a}nchez-Conde}, M.~A., {Palomares-Ruiz}, S., \&
  {Prada}, F. 2017, \mnras, 466, 4974, \dodoi{10.1093/mnras/stx026}

\bibitem[{{Moskowitz} \& {Walker}(2020)}]{MoskowitzWalker2020}
{Moskowitz}, A.~G., \& {Walker}, M.~G. 2020, \apj, 892, 27,
  \dodoi{10.3847/1538-4357/ab7459}

\bibitem[{{Navarro} {et~al.}(1996){Navarro}, {Frenk}, \&
  {White}}]{Navarro1996a}
{Navarro}, J.~F., {Frenk}, C.~S., \& {White}, S. D.~M. 1996, \apj, 462, 563,
  \dodoi{10.1086/177173}

\bibitem[{{Navarro} {et~al.}(1997){Navarro}, {Frenk}, \& {White}}]{Navarro1997}
{Navarro}, J.~F., {Frenk}, C.~S., \& {White}, S.~D.~M. 1997, ApJ, 490, 493,
  \dodoi{10.1086/304888}

\bibitem[{{Navarro} {et~al.}(2004){Navarro}, {Hayashi}, {Power}, {Jenkins},
  {Frenk}, {White}, {Springel}, {Stadel}, \& {Quinn}}]{Navarro2004}
{Navarro}, J.~F., {Hayashi}, E., {Power}, C., {et~al.} 2004, \mnras, 349, 1039,
  \dodoi{10.1111/j.1365-2966.2004.07586.x}

\bibitem[{{Navarro} {et~al.}(2010){Navarro}, {Ludlow}, {Springel}, {Wang},
  {Vogelsberger}, {White}, {Jenkins}, {Frenk}, \& {Helmi}}]{NavarroLudlow2010}
{Navarro}, J.~F., {Ludlow}, A., {Springel}, V., {et~al.} 2010, \mnras, 402, 21,
  \dodoi{10.1111/j.1365-2966.2009.15878.x}

\bibitem[{{Newton} {et~al.}(2018){Newton}, {Cautun}, {Jenkins}, {Frenk}, \&
  {Helly}}]{Newton2018}
{Newton}, O., {Cautun}, M., {Jenkins}, A., {Frenk}, C.~S., \& {Helly}, J.~C.
  2018, \mnras, 479, 2853, \dodoi{10.1093/mnras/sty1085}

\bibitem[{{O{\~n}orbe} {et~al.}(2015){O{\~n}orbe}, {Boylan-Kolchin}, {Bullock},
  {Hopkins}, {Kere{\v{s}}}, {Faucher-Gigu{\`e}re}, {Quataert}, \&
  {Murray}}]{Onorbe2015}
{O{\~n}orbe}, J., {Boylan-Kolchin}, M., {Bullock}, J.~S., {et~al.} 2015,
  \mnras, 454, 2092, \dodoi{10.1093/mnras/stv2072}

\bibitem[{{Ogiya} \& {Hahn}(2018)}]{OgiyaHahn2018}
{Ogiya}, G., \& {Hahn}, O. 2018, \mnras, 473, 4339,
  \dodoi{10.1093/mnras/stx2639}

\bibitem[{{Oman} {et~al.}(2015){Oman}, {Navarro}, {Fattahi}, {Frenk}, {Sawala},
  {White}, {Bower}, {Crain}, {Furlong}, {Schaller}, {Schaye}, \&
  {Theuns}}]{Oman2015}
{Oman}, K.~A., {Navarro}, J.~F., {Fattahi}, A., {et~al.} 2015, \mnras, 452,
  3650, \dodoi{10.1093/mnras/stv1504}

\bibitem[{{Pascale} {et~al.}(2018){Pascale}, {Posti}, {Nipoti}, \&
  {Binney}}]{Pascale2018}
{Pascale}, R., {Posti}, L., {Nipoti}, C., \& {Binney}, J. 2018, \mnras, 480,
  927, \dodoi{10.1093/mnras/sty1860}

\bibitem[{{Pe{\~n}arrubia} \& {Benson}(2005)}]{PenarrubiaBenson2005}
{Pe{\~n}arrubia}, J., \& {Benson}, A.~J. 2005, \mnras, 364, 977,
  \dodoi{10.1111/j.1365-2966.2005.09633.x}

\bibitem[{{Pe{\~n}arrubia} {et~al.}(2010){Pe{\~n}arrubia}, {Benson}, {Walker},
  {Gilmore}, {McConnachie}, \& {Mayer}}]{Penarrubia2010}
{Pe{\~n}arrubia}, J., {Benson}, A.~J., {Walker}, M.~G., {et~al.} 2010, MNRAS,
  406, 1290, \dodoi{10.1111/j.1365-2966.2010.16762.x}

\bibitem[{{Pe{\~n}arrubia} {et~al.}(2008){Pe{\~n}arrubia}, {Navarro}, \&
  {McConnachie}}]{Penarrubia2008}
{Pe{\~n}arrubia}, J., {Navarro}, J.~F., \& {McConnachie}, A.~W. 2008, ApJ, 673,
  226, \dodoi{10.1086/523686}

\bibitem[{{Pe{\~n}arrubia} {et~al.}(2012){Pe{\~n}arrubia}, {Pontzen}, {Walker},
  \& {Koposov}}]{Penarrubia2012}
{Pe{\~n}arrubia}, J., {Pontzen}, A., {Walker}, M.~G., \& {Koposov}, S.~E. 2012,
  \apjl, 759, L42, \dodoi{10.1088/2041-8205/759/2/L42}

\bibitem[{{Pereira-Wilson} {et~al.}(2023){Pereira-Wilson}, {Navarro},
  {Ben{\'\i}tez-Llambay}, \& {Santos-Santos}}]{Pereira-Wilson2023}
{Pereira-Wilson}, M., {Navarro}, J.~F., {Ben{\'\i}tez-Llambay}, A., \&
  {Santos-Santos}, I. 2023, \mnras, 519, 1425, \dodoi{10.1093/mnras/stac3633}

\bibitem[{{Pontzen} \& {Governato}(2012)}]{Pontzen2012}
{Pontzen}, A., \& {Governato}, F. 2012, \mnras, 421, 3464,
  \dodoi{10.1111/j.1365-2966.2012.20571.x}

\bibitem[{{Power} {et~al.}(2003){Power}, {Navarro}, {Jenkins}, {Frenk},
  {White}, {Springel}, {Stadel}, \& {Quinn}}]{Power2003}
{Power}, C., {Navarro}, J.~F., {Jenkins}, A., {et~al.} 2003, \mnras, 338, 14,
  \dodoi{10.1046/j.1365-8711.2003.05925.x}

\bibitem[{{Read} {et~al.}(2018){Read}, {Walker}, \& {Steger}}]{Read2018}
{Read}, J.~I., {Walker}, M.~G., \& {Steger}, P. 2018, \mnras, 481, 860,
  \dodoi{10.1093/mnras/sty2286}

\bibitem[{{Read} {et~al.}(2019){Read}, {Walker}, \& {Steger}}]{Read2019}
---. 2019, \mnras, 484, 1401, \dodoi{10.1093/mnras/sty3404}

\bibitem[{{Riley} {et~al.}(2019){Riley}, {Fattahi}, {Pace}, {Strigari},
  {Frenk}, {G{\'o}mez}, {Grand}, {Marinacci}, {Navarro}, {Pakmor}, {Simpson},
  \& {White}}]{Riley2019}
{Riley}, A.~H., {Fattahi}, A., {Pace}, A.~B., {et~al.} 2019, \mnras, 486, 2679,
  \dodoi{10.1093/mnras/stz973}

\bibitem[{{Santos-Santos} {et~al.}(2020){Santos-Santos}, {Navarro},
  {Robertson}, {Ben{\'\i}tez-Llambay}, {Oman}, {Lovell}, {Frenk}, {Ludlow},
  {Fattahi}, \& {Ritz}}]{Santos-Santos2020}
{Santos-Santos}, I. M.~E., {Navarro}, J.~F., {Robertson}, A., {et~al.} 2020,
  \mnras, 495, 58, \dodoi{10.1093/mnras/staa1072}

\bibitem[{{Schaye} {et~al.}(2015){Schaye}, {Crain}, {Bower}, {Furlong},
  {Schaller}, {Theuns}, {Dalla Vecchia}, {Frenk}, {McCarthy}, {Helly},
  {Jenkins}, {Rosas-Guevara}, {White}, {Baes}, {Booth}, {Camps}, {Navarro},
  {Qu}, {Rahmati}, {Sawala}, {Thomas}, \& {Trayford}}]{Schaye2015}
{Schaye}, J., {Crain}, R.~A., {Bower}, R.~G., {et~al.} 2015, \mnras, 446, 521,
  \dodoi{10.1093/mnras/stu2058}

\bibitem[{{Sestito} {et~al.}(2023){Sestito}, {Zaremba}, {Venn}, {D'Aoust},
  {Hayes}, {Jensen}, {Navarro}, {Jablonka}, {Fern{\'a}ndez-Alvar}, {Glover},
  {McConnachie}, \& {Chen{\'e}}}]{Sestito2023}
{Sestito}, F., {Zaremba}, D., {Venn}, K.~A., {et~al.} 2023, \mnras, 525, 2875,
  \dodoi{10.1093/mnras/stad2427}

\bibitem[{{Simon}(2019)}]{Simon2019Review}
{Simon}, J.~D. 2019, \araa, 57, 375,
  \dodoi{10.1146/annurev-astro-091918-104453}

\bibitem[{{Smith} {et~al.}(2024){Smith}, {Cerny}, {Hayes}, {Sestito}, {Jensen},
  {McConnachie}, {Geha}, {Navarro}, {Li}, {Cuillandre}, {Errani}, {Chambers},
  {Gwyn}, {Hammer}, {Hudson}, {Magnier}, \& {Martin}}]{Smith2023}
{Smith}, S. E.~T., {Cerny}, W., {Hayes}, C.~R., {et~al.} 2024, \apj, 961, 92,
  \dodoi{10.3847/1538-4357/ad0d9f}

\bibitem[{{Spergel} \& {Steinhardt}(2000)}]{Spergel2000}
{Spergel}, D.~N., \& {Steinhardt}, P.~J. 2000, Physical Review Letters, 84,
  3760, \dodoi{10.1103/PhysRevLett.84.3760}

\bibitem[{{Springel} {et~al.}(2008){Springel}, {Wang}, {Vogelsberger},
  {Ludlow}, {Jenkins}, {Helmi}, {Navarro}, {Frenk}, \& {White}}]{Springel2008}
{Springel}, V., {Wang}, J., {Vogelsberger}, M., {et~al.} 2008, MNRAS, 391,
  1685, \dodoi{10.1111/j.1365-2966.2008.14066.x}

\bibitem[{{St{\"u}cker} {et~al.}(2023){St{\"u}cker}, {Ogiya}, {Angulo},
  {Aguirre-Santaella}, \& {S{\'a}nchez-Conde}}]{Stuecker2023}
{St{\"u}cker}, J., {Ogiya}, G., {Angulo}, R.~E., {Aguirre-Santaella}, A., \&
  {S{\'a}nchez-Conde}, M.~A. 2023, \mnras, 521, 4432,
  \dodoi{10.1093/mnras/stad844}

\bibitem[{{Taibi} {et~al.}(2020){Taibi}, {Battaglia}, {Rejkuba}, {Leaman},
  {Kacharov}, {Iorio}, {Jablonka}, \& {Zoccali}}]{Taibi2020}
{Taibi}, S., {Battaglia}, G., {Rejkuba}, M., {et~al.} 2020, \aap, 635, A152,
  \dodoi{10.1051/0004-6361/201937240}

\bibitem[{{Tegmark} {et~al.}(1997){Tegmark}, {Silk}, {Rees}, {Blanchard},
  {Abel}, \& {Palla}}]{Tegmark1997}
{Tegmark}, M., {Silk}, J., {Rees}, M.~J., {et~al.} 1997, \apj, 474, 1,
  \dodoi{10.1086/303434}

\bibitem[{{Tollerud} {et~al.}(2008){Tollerud}, {Bullock}, {Strigari}, \&
  {Willman}}]{Tolerud2008}
{Tollerud}, E.~J., {Bullock}, J.~S., {Strigari}, L.~E., \& {Willman}, B. 2008,
  \apj, 688, 277, \dodoi{10.1086/592102}

\bibitem[{{Torrealba} {et~al.}(2019){Torrealba}, {Belokurov}, \&
  {Koposov}}]{Torrealba2019clusters}
{Torrealba}, G., {Belokurov}, V., \& {Koposov}, S.~E. 2019, \mnras, 484, 2181,
  \dodoi{10.1093/mnras/stz071}

\bibitem[{{Torrealba} {et~al.}(2016){Torrealba}, {Koposov}, {Belokurov}, \&
  {Irwin}}]{Torrealba2016}
{Torrealba}, G., {Koposov}, S.~E., {Belokurov}, V., \& {Irwin}, M. 2016,
  \mnras, 459, 2370, \dodoi{10.1093/mnras/stw733}

\bibitem[{{van den Bosch} \& {Ogiya}(2018)}]{vdBOgiya2018}
{van den Bosch}, F.~C., \& {Ogiya}, G. 2018, \mnras, 475, 4066,
  \dodoi{10.1093/mnras/sty084}

\bibitem[{{van den Bosch} {et~al.}(2018){van den Bosch}, {Ogiya}, {Hahn}, \&
  {Burkert}}]{vdb2018}
{van den Bosch}, F.~C., {Ogiya}, G., {Hahn}, O., \& {Burkert}, A. 2018, \mnras,
  474, 3043, \dodoi{10.1093/mnras/stx2956}

\bibitem[{{van den Bosch} {et~al.}(2005){van den Bosch}, {Tormen}, \&
  {Giocoli}}]{vdBTormen2005}
{van den Bosch}, F.~C., {Tormen}, G., \& {Giocoli}, C. 2005, \mnras, 359, 1029,
  \dodoi{10.1111/j.1365-2966.2005.08964.x}

\bibitem[{{Walker} {et~al.}(2007){Walker}, {Mateo}, {Olszewski}, {Gnedin},
  {Wang}, {Sen}, \& {Woodroofe}}]{Walker2007bLetter}
{Walker}, M.~G., {Mateo}, M., {Olszewski}, E.~W., {et~al.} 2007, \apjl, 667,
  L53, \dodoi{10.1086/521998}

\bibitem[{{Walker} \& {Pe{\~n}arrubia}(2011)}]{Walker2011}
{Walker}, M.~G., \& {Pe{\~n}arrubia}, J. 2011, ApJ, 742, 20,
  \dodoi{10.1088/0004-637X/742/1/20}

\bibitem[{{Wang} {et~al.}(2020){Wang}, {Bose}, {Frenk}, {Gao}, {Jenkins},
  {Springel}, \& {White}}]{Wang2020}
{Wang}, J., {Bose}, S., {Frenk}, C.~S., {et~al.} 2020, \nat, 585, 39,
  \dodoi{10.1038/s41586-020-2642-9}

\bibitem[{{Wang} {et~al.}(2019){Wang}, {de Boer}, {Pieres}, {Li},
  {Drlica-Wagner}, {Koposov}, {Vivas}, {Pace}, {Santiago}, {Walker}, {Tucker},
  {Strigari}, {Marshall}, {Yanny}, {DePoy}, {Bechtol}, {Roodman}, {Abbott},
  {Abdalla}, {Allam}, {Annis}, {Avila}, {Bertin}, {Brooks}, {Burke}, {Carnero
  Rosell}, {Carrasco Kind}, {Cunha}, {D'Andrea}, {da Costa}, {De Vicente},
  {Desai}, {Eifler}, {Estrada}, {Flaugher}, {Frieman}, {Garc{\'\i}a-Bellido},
  {Gerdes}, {Gruen}, {Gruendl}, {Gutierrez}, {Hollowood}, {Honscheid}, {James},
  {Kuehn}, {Kuropatkin}, {Lahav}, {Maia}, {Miquel}, {Sanchez}, {Scarpine},
  {Sevilla-Noarbe}, {Smith}, {Smith}, {Sobreira}, {Suchyta}, {Swanson},
  {Tarle}, \& {DES Collaboration}}]{Wang2019}
{Wang}, M.~Y., {de Boer}, T., {Pieres}, A., {et~al.} 2019, \apj, 881, 118,
  \dodoi{10.3847/1538-4357/ab31a9}

\bibitem[{{White} \& {Rees}(1978)}]{WhiteRees1978}
{White}, S.~D.~M., \& {Rees}, M.~J. 1978, \mnras, 183, 341,
  \dodoi{10.1093/mnras/183.3.341}

\bibitem[{{Woo} {et~al.}(2008){Woo}, {Courteau}, \& {Dekel}}]{Woo2008}
{Woo}, J., {Courteau}, S., \& {Dekel}, A. 2008, \mnras, 390, 1453,
  \dodoi{10.1111/j.1365-2966.2008.13770.x}

\bibitem[{{Zavala} \& {Frenk}(2019)}]{ZavalaFrenk2019}
{Zavala}, J., \& {Frenk}, C.~S. 2019, Galaxies, 7, 81,
  \dodoi{10.3390/galaxies7040081}

\end{thebibliography}

\label{lastpage}
\end{document}